\documentclass[runningheads]{llncs}
\usepackage[table]{xcolor}

\usepackage{graphicx}
\graphicspath{{./figures/}}

\usepackage[misc]{ifsym}
\usepackage{array,multirow,tikz,wrapfig}
\usetikzlibrary{shapes.geometric, arrows}
\usepackage{empheq}
\usepackage[most]{tcolorbox}
\newtcbox{\mymath}[1][]{%
    nobeforeafter, math upper, tcbox raise base,
    enhanced, colframe=blue!30!black,
    colback=blue!30, boxrule=1pt,
    #1}

\tikzset{base/.style={draw, align=center, minimum height=4ex},
         test1/.style={base, diamond, aspect=2, text width=5em, inner sep=5pt},
         test2/.style={base, diamond, aspect=2, text width=5em, inner sep=-4.8pt}
        }  

\usepackage[ruled]{algorithm2e}
\usepackage{amsmath,amssymb}
\usepackage{xspace}
\usepackage{url}
\usepackage{bibnames}
\usepackage{subfig}
\usepackage{wrapfig}

\usepackage{tikz}
\usetikzlibrary{arrows}
\usetikzlibrary{external}
\tikzset{vertex/.style = {shape=circle,draw,minimum size=1.5em}}
\tikzset{edge/.style = {->,> = latex'}}

\begin{document}
\title{TwitterMancer: \\
Predicting Interactions on Twitter Accurately}
%
%
\author{Konstantinos Sotiropoulos \inst{1} \and
John W. Byers \inst{1}  \and
Polyvios Pratikakis \inst{2} \and Charalampos E. Tsourakakis \inst{1}}
\authorrunning{A preprint}
%
\institute{Boston University, Boston MA, USA \and
University of Crete, Heraklion Crete, Greece}
\maketitle         

\makeatletter
\newcommand*{\rom}[1]{\expandafter\@slowromancap\romannumeral #1@}
\makeatother

\newcommand{\Oh}{\mathcal{O}}
\newcommand{\maj}{\mathrm{maj}}
\newcommand{\hide}[1]{} 

\newcommand{\ber}[1]{\text{Bernoulli}\left(#1\right)}
\newcommand{\bin}[2]{\text{Bin}\left(#1,#2\right)}
\newcommand{\geom}[1]{\text{Geom}\left(#1\right)}
\newcommand{\hypg}[3]{\text{Hypergeometric}\left(#1,#2,#3\right)}
\newcommand{\unif}[1]{\text{Uniform}\left(#1\right)}
\newcommand{\KL}[2]{D_{\text{KL}}({#1}||{#2} )}

\newcommand{\pr}[1]{\ensuremath{{\bf{Pr}}\left[{#1}\right]}}

\newcommand{\cc}{\text{Correlation Clustering}\xspace}
\newcommand{\cccc}{\text{2-Correlation-Clustering}\xspace}
\newcommand{\f}{\tilde{f}}
\def\e{\epsilon}
\def\hT{\widehat{T}}
 \def\r{\rho}
\newcommand{\diam}{\frac{ \log{n}}{\log{\log{n}}}}
\def\hD{\widehat{D}}
\def\g{\gamma}
\def\risk{risk}
\newcommand{\beql}[1]{\begin{equation}\label{#1}}

\newcommand{\beq}[1]{\begin{equation}\label{#1}}
\newcommand{\eeq}{\end{equation}}
\newcommand{\sfrac}[2]{\frac{\scriptstyle #1}{\scriptstyle #2}}
\newcommand{\bfrac}[2]{\left(\frac{#1}{#2}\right)}
\newcommand{\brac}[1]{\left(#1\right)}
\def\a{\alpha}
\newcommand{\misc}[1]{{\color{magenta}#1}}
\newcommand{\reminder}[1]{{\color{red}#1}}
 \newcommand{\vectornorm}[1]{\left|\left|#1\right|\right|}
\newcommand{\field}[1]{\mathbb{#1}} 
\newcommand{\One}[1]{\ensuremath{{\mathbf 1}\left(#1\right)}}
\newcommand{\Prob}[1]{\ensuremath{{\bf{Pr}}\left[{#1}\right]}}
\newcommand{\Mean}[1]{\ensuremath{{\mathbb E}\left[{#1}\right]}}
\newcommand{\NP}{\ensuremath{\mathbf{NP}}\xspace}
\newcommand{\NPhard}{{\ensuremath{\mathbf{NP}}-hard}\xspace}
\newcommand{\NPcomplete}{{\ensuremath{\mathbf{NP}}-complete}\xspace}
\newcommand{\sgn}{{\ensuremath{\mathrm{sgn}}}}
\newcommand{\whp}{\textit{whp}\xspace}
\newcommand{\Var}[1]{{\mathbb Var}\left[{#1}\right]}

\newcommand{\spara}[1]{\smallskip\noindent{\bf #1}}
\newcommand{\mpara}[1]{\medskip\noindent{\bf #1}}
\newcommand{\para}[1]{\noindent{\bf #1}}

\newcommand{\pairfeature}[3]{ \ensuremath{P_{#1}(#2, #3)} }
\newcommand{\bluedot}{\textcolor{blue!80!black}{\bullet}}
\newcommand{\reddotdot}{\textcolor{red!80!black}{\scriptstyle\bullet\bullet}}

\begin{abstract} 
This paper investigates the interplay between different types of
user interactions on Twitter, with respect to predicting missing or
unseen interactions.

For example, given a set of {\em retweet} interactions between Twitter
users, how accurately can we predict {\em reply} interactions? Is it
more difficult to predict {\em retweet} or {\em quote} interactions
between a pair of accounts?  Also, how important is time locality, and
which features of interaction patterns are most important to enable
accurate prediction of specific Twitter interactions?

Our empirical study of Twitter interactions contributes initial answers to these questions.

We have crawled an extensive dataset of  Greek-speaking Twitter
accounts and their {\em follow, quote, retweet, reply} interactions
over a period of a month.  

We find we can accurately predict many interactions of Twitter users.
Interestingly, the most predictive features vary with the user profiles,
and are not the same across all users.

For example, for a pair of users that interact with a large number of
other Twitter users, we find that certain ``higher-dimensional''
triads, i.e., triads that involve multiple types of interactions, are
very informative, whereas for less active Twitter users, certain in-degrees
and out-degrees play a major role.  
Finally, we provide various other insights on Twitter user behavior.  

Our code and data are available at \url{https://github.com/twittermancer/}. 

\end{abstract}

\section{Introduction} 
\label{sec:intro} 

Twitter is a microblogging service with more than 300 million monthly
active users worldwide, as of early 2018. 
Its unique characteristics have drawn the attention of many
researchers, and provide a novel opportunity for understanding human
behavior in a public and observable forum at an unprecedented scale. 
Among many other applications, the Twitter repository of human signals
has been used to predict 
the stock market \cite{bollen2011twitter}, estimate mortality of heart
diseases \cite{eichstaedt2015psychological}, forecast election
outcomes \cite{hurlimann2016twitter,enli2017twitter}, and detect
humanitarian crises in real time~\cite{mendoza2010twitter,sakaki2010earthquake}. 
Twitter follows the pulse of global society, and therefore
studying it from all possible angles is an active area of research. 
The angle we take here is related to the multiple networks that
naturally underlie the Twitter platform. Specifically, over a fixed
window of observation, a user $u$ can interact with a user $v$ in more
than one way;  $u$ may follow, reply, quote, retweet $v$, or like a
tweet of $v$, or send her a message. These types of interactions
naturally define a {\em multilayer} directed network, with layers
corresponding to the types of interactions that occur.  
We study both the unweighted and weighted version of these networks,
but for simplicity, we discuss the unweighted (0/1) case herein.
In this setting, the sets of adjacencies (directed edges) across
layers are typically correlated, but also exhibit clear differences.
We are interested in the extent to which those differences can be
characterized, i.e., differences that are specific to certain layers
(e.g., sparsity of a given type of interaction), and differences
that relate to the local neighborhoods of users (e.g., graph structure
around a celebrity).
Difference characterization via relevant feature analysis enables
accurate cross-layer prediction;  conversely, differences that are
hard to characterize makes cross-layer prediction more difficult.
In this work we focus on the following question: 

\begin{tcolorbox}
\begin{problem}
\label{question}  
Given two Twitter users $u,v$, and the Twitter multilayer network graph, can we predict what type of interactions will take place between $u,v$?
\end{problem}
\end{tcolorbox}

For the purpose of answering Problem~\ref{question}, we have crawled a
large Twitter dataset of Greek-speaking users, spanning the full month
of February 2018.
Using this corpus of tweets, we have created a multilayer network with
four different layers, correspond to four different types of
interactions respectively: {\em follow, quote, reply, retweet}.
We use this network, together with the temporal information available
to us, to attack a series of problems that we summarize in the
following.

\spara{High-dimensional link prediction.}
We formulate the Twitter link prediction problem as follows
Suppose we are given the multilayer Twitter network, except for
all interactions between a {\em pair} $u,v$ of nodes that are known to have
interacted.
How reliably can we infer whether $u,v$ will {\em follow, reply,
quote}, or {\em retweet} each other using the information provided by
the rest of the network?  Our approach is data-driven, generalizes
the seminal work of Liben-Nowell and Kleinberg~\cite{liben2007link} on
link prediction, and follows the established framework of
Leskovec, Kleinberg and Huttenlocher \cite{leskovec2010predicting}.
Our main finding is that leveraging information from other types of
interactions boosts prediction accuracy significantly, on the order
of 9--32\%. 
We observe that typically higher gains are obtained for the sparser
interaction layers, and that overall the simplest forms of
interaction, like retweeting (just two clicks), are easiest to predict.

\spara{Temporal aspect.}
We perform the classification experiments on a daily basis spanning
the period of a month (February 2018). We observe that the prediction
accuracy remains stable over time for each given type of interaction
(i.e., there is no day-of-week effect), but it does range
significantly across different interaction types. 

\spara{Correlation between types of interactions.}
We perform a detailed study on how certain interactions increase or
decrease the likelihood of other interactions. We focus on two types
of experiments to assess these interaction correlations. First, 
for each interaction, we test how the prediction accuracy changes for the
standard link prediction formulation when we use information from one 
additional layer.  
Figure~\ref{fig:highlight} displays a heatmap depicting the classification
accuracies of a specific $(u,v)$ interaction (row) by leveraging other
interactions of that type plus an additional interaction (column).  
Entries along the diagonal correspond to using no outside interaction types.
For example, consider the Reply row.  When only Replies are used in
prediction, the prediction accuracy is a relatively low 58.1\% on the
diagonal  
(as described in Section~\ref{sec:framework}, 
we test against an equal number of edges and non-edges so that 50\% is achieved by random guessing).  
The prediction accuracy is boosted to 75.5\% using the additional information from
Retweets (rightmost column), but only to 63.4\% using Quotes (second column). 
We observe that for all interactions, side information is always beneficial,
as one would anticipate.  The most informative other interaction is
consistently Retweets, and for Retweets themselves, the most informative
interaction is from Quotes (even though that graph layer is significantly 
sparser than, for example, the follows layer).
Finally, our interpretable logistic regression model, presented in Section~\ref{sec:framework},
allows us to understand the relative importance of certain types of
interactions.

\begin{figure*}[t]
\centering
\includegraphics[scale=0.55]{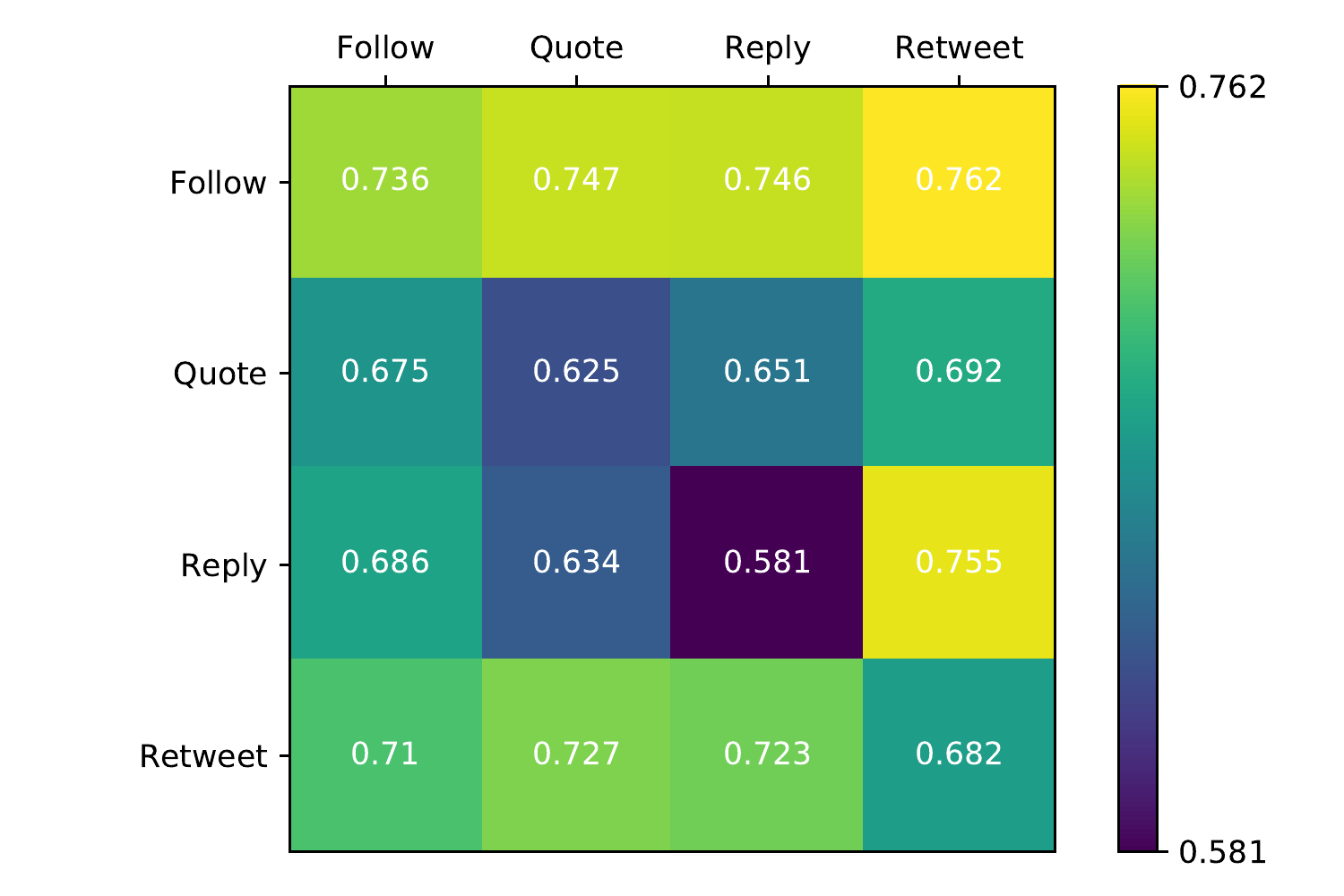}
\caption{Predict edges of a type on the horizontal axis, using also
information from another type of the vertical axis. Diagonal tiles
refer to using only one type to predict this type.}
\label{fig:highlight}
\end{figure*}

\spara{Hardness of prediction as a function of user profiles.}
Finally, we provide novel insights on what features are important for
accurate prediction for different user profiles. Our findings
strongly suggest a dichotomy between two types of pairs of users;
pairs of users that engage jointly with many other Twitter users in
various combinations of interactions,  and those who don't (for
details see Section~\ref{sec:results}). For the former, features
involving triads turn out to be important features in accurate prediction, 
complementary to degree-based features that are useful 
for less well-connected users.
We provide a detailed study of how 
all these features jointly affect the
empirical effectiveness
of link prediction. 
Our findings can be seen as a higher dimensional
analog of the importance of triads in the signed link prediction
problem~\cite{leskovec2010predicting}.

\spara{Roadmap.}
Section~\ref{sec:related} presents closely related work.
Sections~\ref{sec:framework},~\ref{sec:dataset}, and~\ref{sec:results}
present our machine learning framework, discuss the dataset
collection methods we use, and our experimental findings, respectively.
Section~\ref{sec:concl} concludes our paper.

\section{Related Work}
\label{sec:related}

Due to the large volume of work related to Twitter and link
prediction, we focus on 
 related literature that lies closest to our work.

\spara{Link prediction.}
The link prediction problem was popularized by the seminal paper of
Libel-Nowell and Kleinberg~\cite{liben2007link}. Since then, link
prediction has been studied extensively~\cite{al2011survey}.
Close to our work lies the framework proposed by Leskovec et
al.~\cite{leskovec2010predicting} that extended the link prediction
problem to graphs with positive and negative interactions between
nodes.  Their work suggests a machine learning framework that uses
local features and a logistic regression classifier to predict the
unknown sign of an edge.

\spara{Prediction on Twitter.}
A wide variety of prediction problems have been studied on Twitter, due
to its unique nature.
Petrovic, Osborne and Lavrenko \cite{petrovic2011rt} studied the
problem of predicting whether a user will retweet a particular tweet,
or more generally spread an item of interest.
On the same problem, Galuba et al.~\cite{galuba2010outtweeting} used a
propagation model to find which users are likely to mention certain
URLs.
Martin{\v{c}}i{\'c}-Ip{\v{s}}i{\'c} et al.~\cite{martinvcic2017link}
focus on predicting pairs of hashtags (or words) that will co-occur in
future tweets.

Jalili et al.~\cite{jalili2017link} focused on the following link
prediction problem: given a set of users who participate both on
Twitter and Foursquare, predict links between users at Foursquare by
using information from the Twitter network. While experimentally they
do not study the inverse questions, their tools can be used to predict
links on Twitter from Foursquare.
Hristova et al.~\cite{hristova2016multilayer} enrich this framework by
the use of random forest classifiers. 
 
Our work is however, the first ---to the best of our knowledge---
that explores link prediction in the context of different interactions
among users on the Twitter network. Abufouda and
Zweig~\cite{abufouda2015we} use multiple networks to predict which
links among users represent actual, real-life links.  Our work differs
from the bulk of such Twitter-related link prediction problems as 
we focus on predicting interactions on, and across different Twitter
layers.

\section{Proposed Framework}
\label{sec:framework}

In this work we focus on the following question that extends the
seminal formulation of Leskovec, Kleinberg, and
Huttenlocher~\cite{leskovec2010predicting} on predicting signed edges
in online social networks, and more generally research work on link
prediction~\cite{liben2007link}:

\begin{tcolorbox}
\begin{problem}
\label{prob_twitter}  
Given the Twitter graph  containing user accounts and their pairwise
{\em follow, reply, retweet}, and {\em quote} interactions, and a pair
of user accounts $\{u,v\}$, how accurately can we predict whether $u$
will follow, reply, retweet, or quote $v$? 
\end{problem}
\end{tcolorbox}

Understanding Problem~\ref{prob_twitter} will contribute further
towards a better understanding of user behavior on Twitter, and may
lead to detecting correlations between types of interactions that will
be useful for anomaly detection among others. We model the input
dataset as a directed, multi-label, multigraph $G = (V, E, I, \ell_E)$.
Specifically, the node and edge sets $V,E$ correspond to the set of
Twitter user accounts, and the interactions among them, respectively.
Different types of interactions are modeled by the label function,
i.e., $\ell_E:E\to \mathcal{I}$ is the function that labels each edge
according to the set $\mathcal{I}$ of all possible interactions. Here,
we consider $\mathcal{I}=\{\text{follow, quote, reply, retweet}\}$. Our framework naturally extends to larger sets of
interactions and also weighted graphs, i.e., graphs where each edge
is associated with the counts of interactions.
 
\subsection{A Machine Learning Framework}
The task of predicting a missing edge on a graph can be thought of as
the following classification problem: given a pair of users $(u,v)$
and an interaction type $i$, we are trying to learn a function $f$
that returns 1 if an edge $(u,v)$ with label $i$ is present on the
graph, and -1 otherwise.
To tackle our problem, while retaining interpretability of results, we
use a simple logistic regression framework. We use the term {\em
embeddedness} ---as used also by Leskovec et al.~\cite{leskovec2010predicting}---
for an edge $(u,v)$ as the quantity 
$| \{t |(u, t) \mbox{ and } (t, v) \in E\}|$, the number of common
neighbors between $\{u,v\}$.

\spara{Features:} As Twitter graphs are typically on the order of 
millions of users, we
use local features that are computationally efficient to extract. 
This approach also mirrors a local view that Twitter users usually
have (e.g., on their timeline) when deciding to make an action (follow
another user, reply to a tweet, etc.). 
We build on features already used in relevant related work, while also
incorporating new feature sets that capture the interplay between
different types of interactions.

The first set of features that we use aims at capturing the 
propensity of users to interact with other users more or less often. 
To capture the breadth and relative frequency of activity of each
user, we define features based on the degrees of corresponding nodes
in the interaction graph.
As Twitter is inherently a directed network, and since we are concerned with
inferring the directionality of interaction $(u,v)$, we use the
following 10 directed \textit{degree features}.
We use the \textit{out-degree} of user $u$ for each of the interaction
types in $\mathcal{I}$ (4 features),
the \textit{in-degree} of user $v$ for those types (4 features), as
well as counts of the {{\em number}} of different interaction types
$u$ initiated and $v$ received, respectively (2 features). 

The second set of features we use considers a common neighbor $t$ of
$u$ and $v$ (as counted when computing the embeddedness of $(u, v)$),
and identifies all possible ways in which $t$ had an
interaction with both $u$ and $v$. 
For this set of features, we consider each interaction type separately
and retain the directionality of edges. 
We have, thus, $3 \times 3$ possible triads for each type, times 4
interaction types, yielding 36 features in total (see
Fig.~\ref{single_triads}).

For the final set of features, we again employ triads as above, but
this time we make use of the interplay between different types of
interactions, e.g., for a common neighbor $t$, if $v$ retweets
something that $t$ posted, and $u$ replies on $t$, then it is likely
that $u$ follows $v$.  To keep the cardinality of this set of features
manageable and to avoid overfitting, we drop the directionality of the
edges and use pairs of different interactions. 
Thus, we have again $3\times 3$ different triads for each one of the
$\binom{4}{2}$ pairs, yielding 54 additional features (see
Fig.~\ref{pairwise_triads}). 

As our predictions are directed ($u\to v$), we will use the following
notation for the feature names whenever we refer to them: {\em Out(i)} will
refer to the out-degree of $u$ at layer (interaction) $i$ and respectively {\em In(i)} for
the in-degree of $v$. We will use the first letter of interactions\footnote{We distinguish between reply and
retweets, by using {\em r } and {\em rt }, respectively.} to refer to every layer $i$.
{\em Total(u)} and {\em Total(v)} will refer to the number of different layers
$u$ was an initiator and $v$ a receiver. Lastly, the notation for the triadic features is the one we describe in
Figures \ref{single_triads} and \ref{pairwise_triads}.

\spara{Methodology.}
Using these 100 features, we train a logistic regression model of
the form:
\begin{equation}
\Pr[e\in E|x] = \frac{1}{1+e^{-(b + <w,x>)}}
\end{equation}
where $x$ is a vector representing the 100 features for a sample,
while $b$ is the intercept and $w$ is the vector of coefficients that 
we want to learn.

For each type of interaction (e.g., reply), we randomly sample an
equal number of edges where there was an interaction of this type and
where there was not. 
Therefore, our datasets will be balanced, providing a baseline of how
much more accurately we can predict over random guessing.
We use 10-fold cross-validation: 10 disjoint folds, where within each
fold, $90\%$ of the edges will be used for training and the remaining
$10\%$ for validation.

\begin{figure}[t]
  \centering
  \resizebox{1.0\textwidth}{!}{
\begin{minipage}[b]{.11\textwidth}
\begin{tikzpicture}
\node[vertex, scale=0.8] (1) at (0,0) {$u$};
\node[vertex, scale=0.8] (2) at (1.0,0) {$v$};
\node[vertex, scale=0.8] (3) at (0.5,0.7) {$t$};
\draw[edge] (3) to[bend left=0] (1);
\draw[edge] (3)  to[bend left=0] (2);
\node [below=1cm,text centered] at (3) { $T_1(i)$ };
\end{tikzpicture}
\end{minipage}
\,
\begin{minipage}[b]{.11\textwidth}
\begin{tikzpicture}
\tikzset{vertex/.style = {shape=circle,draw,minimum size=1.5em}}
\tikzset{edge/.style = {->,> = latex'}}
\node[vertex, scale=0.8] (1) at (0,0) {$u$};
\node[vertex, scale=0.8] (2) at (1.0,0) {$v$};
\node[vertex, scale=0.8] (3) at (0.5,0.7) {$t$};
\draw[edge] (1)  to[bend left=0] (3);
\draw[edge] (2) to[bend left=0] (3);
\node [below=1cm,text centered] at (3) { $T_2(i)$ };
\end{tikzpicture}
\end{minipage}
\,
\begin{minipage}[b]{.11\textwidth}
\begin{tikzpicture}
\node[vertex, scale=0.8] (1) at (0,0) {$u$};
\node[vertex, scale=0.8] (2) at (1.0,0) {$v$};
\node[vertex, scale=0.8] (3) at (0.5,0.7) {$t$};
\draw[edge] (1)  to[bend left=0] (3);
\draw[edge] (3)  to[bend left=0] (2);
\node [below=1cm,text centered] at (3) { $T_3(i)$ };
\end{tikzpicture}
\end{minipage}
\,
\begin{minipage}[b]{.11\textwidth}
\begin{tikzpicture}
\node[vertex, scale=0.8] (1) at (0,0) {$u$};
\node[vertex, scale=0.8] (2) at (1.0,0) {$v$};
\node[vertex, scale=0.8] (3) at (0.5,0.7) {$t$};
\draw[edge] (3) to[bend left=0] (1);
\draw[edge] (2) to[bend left=0] (3);
\node [below=1cm,text centered] at (3) { $T_4(i)$ };
\end{tikzpicture}
\end{minipage}
\,
\begin{minipage}[b]{.11\textwidth}
\begin{tikzpicture}
\node[vertex, scale=0.8] (1) at (0,0) {$u$};
\node[vertex, scale=0.8] (2) at (1.0,0) {$v$};
\node[vertex, scale=0.8] (3) at (0.5,0.7) {$t$};
\draw[edge] (1)  to[bend left=0] (3);
\draw[edge] (3)  to[bend left=10] (2);
\draw[edge] (2) to[bend left=10] (3);
\node [below=1cm,text centered] at (3) { $T_5(i)$ };
\end{tikzpicture}
\end{minipage}
\,
\begin{minipage}[b]{.11\textwidth}
\begin{tikzpicture}
\node[vertex, scale=0.8] (1) at (0,0) {$u$};
\node[vertex, scale=0.8] (2) at (1.0,0) {$v$};
\node[vertex, scale=0.8] (3) at (0.5,0.7) {$t$};
\draw[edge] (1) to[bend left=10] (3);
\draw[edge] (3) to[bend left=10] (1);
\draw[edge] (2) to[bend left=0] (3);
\node [below=1cm,text centered] at (3) { $T_6(i)$ };
\end{tikzpicture}
\end{minipage}
\,
\begin{minipage}[b]{.11\textwidth}
\begin{tikzpicture}
\node[vertex, scale=0.8] (1) at (0,0) {$u$};
\node[vertex, scale=0.8] (2) at (1.0,0) {$v$};
\node[vertex, scale=0.8] (3) at (0.5,0.7) {$t$};
\draw[edge] (3) to[bend left=0] (1);
\draw[edge] (3) to[bend left=10] (2);
\draw[edge] (2) to[bend left=10] (3);
\node [below=1cm,text centered] at (3) { $T_7(i)$ };
\end{tikzpicture}
\end{minipage}
\,
\begin{minipage}[b]{.11\textwidth}
\begin{tikzpicture}
\node[vertex, scale=0.8] (1) at (0,0) {$u$};
\node[vertex, scale=0.8] (2) at (1.0,0) {$v$};
\node[vertex, scale=0.8] (3) at (0.5,0.7) {$t$};
\draw[edge] (1) to[bend left=10] (3);
\draw[edge] (3) to[bend left=10] (1);
\draw[edge] (3) to[bend left=0] (2);
\node [below=1cm,text centered] at (3) { $T_8(i)$ };
\end{tikzpicture}
\end{minipage}
\,
\begin{minipage}[b]{.11\textwidth}
\begin{tikzpicture}
\node[vertex, scale=0.8] (1) at (0,0) {$u$};
\node[vertex, scale=0.8] (2) at (1.0,0) {$v$};
\node[vertex, scale=0.8] (3) at (0.5,0.7) {$t$};
\draw[edge] (1) to[bend left=10] (3);
\draw[edge] (3) to[bend left=10] (1);
\draw[edge] (3) to[bend left=10] (2);
\draw[edge] (2) to[bend left=10] (3);
\node [below=1cm,text centered] at (3) { $T_9(i)$ };
\end{tikzpicture}%
\end{minipage}%
}%
  \caption{Different types of triads involving only one interaction
  type $i \in \mathcal{I}$, the set of all possible interactions,
  while taking into account edge direction.}
  \label{single_triads}
\end{figure}
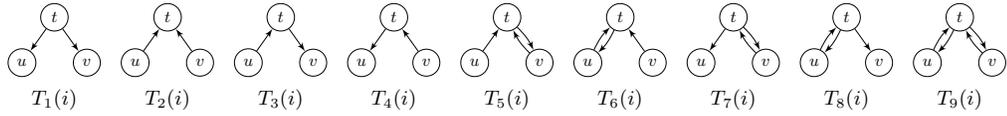

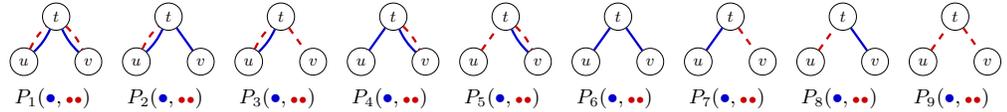
\begin{figure}[t]
  \resizebox{\textwidth}{!}{
\begin{minipage}[b]{.11\textwidth}
\begin{tikzpicture}
\node[vertex, scale=0.8] (1) at (0,0) {$u$};
\node[vertex, scale=0.8] (2) at (1.0,0) {$v$};
\node[vertex, scale=0.8] (3) at (0.5,0.7) {$t$};
\draw[-,red!80!black,line width=1pt] (1) to[bend left=10] (3) [dashed];
\draw[-,blue!80!black,line width=1pt] (1) to[bend left=-10] (3);

\draw[-,red!80!black,line width=1pt] (3) to[bend left=10] (2) [dashed];
\draw[-,blue!80!black,line width=1pt] (3) to[bend left=-10] (2);
\node [below=1cm,text centered] at (3) { \pairfeature{1}{\bluedot}{\reddotdot} };
\end{tikzpicture}
\end{minipage}
\,
\begin{minipage}[b]{.11\textwidth}
\begin{tikzpicture}
\node[vertex, scale=0.8] (1) at (0,0) {$u$};
\node[vertex, scale=0.8] (2) at (1.0,0) {$v$};
\node[vertex, scale=0.8] (3) at (0.5,0.7) {$t$};
\draw[-,red!80!black,line width=1pt] (1) to[bend left=10] (3) [dashed];
\draw[-,blue!80!black,line width=1pt] (1) to[bend left=-10] (3);
\draw[-,blue!80!black,line width=1pt] (3) -- (2);
\node [below=1cm,text centered] at (3) { \pairfeature{2}{\bluedot}{\reddotdot} };
\end{tikzpicture}
\end{minipage}
\,
\begin{minipage}[b]{.11\textwidth}
\begin{tikzpicture}
\node[vertex, scale=0.8] (1) at (0,0) {$u$};
\node[vertex, scale=0.8] (2) at (1.0,0) {$v$};
\node[vertex, scale=0.8] (3) at (0.5,0.7) {$t$};
\draw[-,red!80!black,line width=1pt] (1) to[bend left=10] (3) [dashed];
\draw[-,blue!80!black,line width=1pt] (1) to[bend left=-10] (3);
\draw[-,red!80!black,line width=1pt] (3) -- (2) [dashed];
\node [below=1cm,text centered] at (3) { \pairfeature{3}{\bluedot}{\reddotdot} };
\end{tikzpicture}
\end{minipage}
\,
\begin{minipage}[b]{.11\textwidth}
\begin{tikzpicture}
\node[vertex, scale=0.8] (1) at (0,0) {$u$};
\node[vertex, scale=0.8] (2) at (1.0,0) {$v$};
\node[vertex, scale=0.8] (3) at (0.5,0.7) {$t$};
\draw[-,blue!80!black,line width=1pt] (1) -- (3);
\draw[-,red!80!black,line width=1pt] (3) to[bend left=10] (2) [dashed];
\draw[-,blue!80!black,line width=1pt] (3) to[bend left=-10] (2);
\node [below=1cm,text centered] at (3) { \pairfeature{4}{\bluedot}{\reddotdot} };
\end{tikzpicture}
\end{minipage}
\,
\begin{minipage}[b]{.11\textwidth}
\begin{tikzpicture}
\node[vertex, scale=0.8] (1) at (0,0) {$u$};
\node[vertex, scale=0.8] (2) at (1.0,0) {$v$};
\node[vertex, scale=0.8] (3) at (0.5,0.7) {$t$};
\draw[-,red!80!black,line width=1pt] (1) -- (3) [dashed];
\draw[-,red!80!black,line width=1pt] (3) to[bend left=10] (2) [dashed];
\draw[-,blue!80!black,line width=1pt] (3) to[bend left=-10] (2);
\node [below=1cm,text centered] at (3) { \pairfeature{5}{\bluedot}{\reddotdot} };
\end{tikzpicture}
\end{minipage}
\,
\begin{minipage}[b]{.11\textwidth}
\begin{tikzpicture}
\node[vertex, scale=0.8] (1) at (0,0) {$u$};
\node[vertex, scale=0.8] (2) at (1.0,0) {$v$};
\node[vertex, scale=0.8] (3) at (0.5,0.7) {$t$};
\draw[-,blue!80!black,line width=1pt] (1) to (3);

\draw[-,blue!80!black,line width=1pt] (3) to (2);
\node [below=1cm,text centered] at (3) { \pairfeature{6}{\bluedot}{\reddotdot} };
\end{tikzpicture}
\end{minipage}
\,
\begin{minipage}[b]{.11\textwidth}
\begin{tikzpicture}
\tikzset{vertex/.style = {shape=circle,draw,minimum size=1.5em}}
\tikzset{edge/.style = {->,> = latex'}}
\node[vertex, scale=0.8] (1) at (0,0) {$u$};
\node[vertex, scale=0.8] (2) at (1.0,0) {$v$};
\node[vertex, scale=0.8] (3) at (0.5,0.7) {$t$};
\draw[-,blue!80!black,line width=1pt] (1) -- (3);
\draw[-,red!80!black,line width=1pt] (3) -- (2) [dashed];
\node [below=1cm,text centered] at (3) { \pairfeature{7}{\bluedot}{\reddotdot} };
\end{tikzpicture}
\end{minipage}
\,
\begin{minipage}[b]{.11\textwidth}
\begin{tikzpicture}
\node[vertex, scale=0.8] (1) at (0,0) {$u$};
\node[vertex, scale=0.8] (2) at (1.0,0) {$v$};
\node[vertex, scale=0.8] (3) at (0.5,0.7) {$t$};
\draw[-,red!80!black,line width=1pt] (1) -- (3) [dashed];
\draw[-,blue!80!black,line width=1pt] (3) -- (2);
\node [below=1cm,text centered] at (3) { \pairfeature{8}{\bluedot}{\reddotdot} };
\end{tikzpicture}
\end{minipage}
\,
\begin{minipage}[b]{.11\textwidth}
\begin{tikzpicture}
\node[vertex, scale=0.8] (1) at (0,0) {$u$};
\node[vertex, scale=0.8] (2) at (1.0,0) {$v$};
\node[vertex, scale=0.8] (3) at (0.5,0.7) {$t$};
\draw[-,red!80!black,line width=1pt] (1) -- (3) [dashed];
\draw[-,red!80!black,line width=1pt] (3) -- (2) [dashed];
\node [below=1cm,text centered] at (3) { \pairfeature{9}{\bluedot}{\reddotdot} };
\end{tikzpicture}
\end{minipage}
}%
  \caption{Different types of triads involving the interplay between
  pairs of interactions $\bluedot, \reddotdot \in \mathcal{I}$, the
  set of all possible interactions.
  Whenever a solid or dashed edge is absent, they did not have this type of
  interaction directly.}
  \label{pairwise_triads}
\end{figure}

\section{Dataset collection}
\label{sec:dataset}

\begin{figure}[t]
  \centering
  \begin{minipage}{0.4\linewidth}
  \centering
    \includegraphics[width=\linewidth]{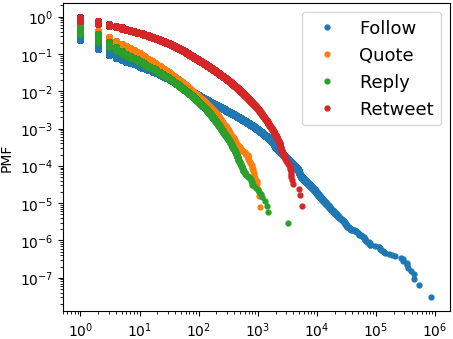}
    (a) Outdegree
  \end{minipage}
  \hfil
  \begin{minipage}{0.4\linewidth}
  \centering
    \includegraphics[width=\linewidth]{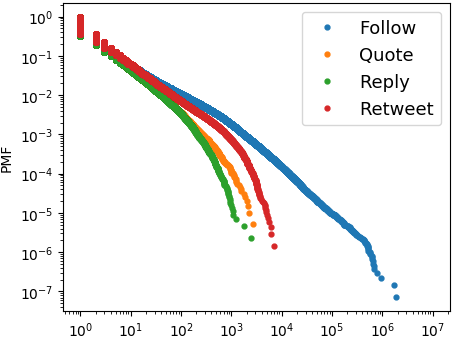}
    (b) Indegree
  \end{minipage}
  \caption{Degree distributions on log-log scale}
  \label{fig4}
\end{figure}

We used an open-source Twitter API crawler to monitor Twitter traffic
generated during February 2018~\cite{pratikakis2018twawler}.  The
crawler targets the Greek-speaking users of Twitter, and performs a
near-total crawl of all tweets by the selected users.  Focusing the
crawler in such a way produces a dense sample of a localized part of
the Twitter graph, instead of a sparse random sample of the whole
graph, as the language ---or alphabet--- barrier facilitates
recognizing interesting users with high probability of locality.  A
similar technique has been applied in the past for the Korean-speaking
part of Twitter~\cite{kwak2011fragile}.

The resulting dataset contains 21 million tweets, of which 9.8 million
are in Greek.  
There are 204 million follow relations among users that were
observed before the start of February 2018, and 33 million additional 
follow relations crawled during February 2018 proper. 
We obtain user information on 13 million unique users, and we classify
340 thousand of them as Greek-speaking, using the conservative rule
of thumb of having posted more than 100 tweets, at least 20\% of which
are in Greek.  The dataset contains many more Greek-speaking users 
than those classified 
marked as Greek-speaking, because (i) many accounts have not yet
posted enough tweets for our heuristic to label them as Greek-speaking 
or not,
and (ii) even if they have shown zero evidence of tweeting in Greek,
they are followed-by or are following Greek speakers.
From this Twitter dataset we extract four graphs, namely Follow,
Quote, Reply, and Retweet, described below.  Figures~\ref{fig4}(a) and
~\ref{fig4}(b) show their 
respective out-degree and in-degree distributions, in log-log scale.

The \textbf{Follow Graph} is the directed graph of the follow relation
among users.  The crawler uses the Twitter API to periodically
scan all tracked users for their lists of friends and followers.
Newly discovered users are given priority in scanning, but 
after the first scan of
friends and followers, users are revisited in a FIFO order that
requires several months to cycle through.  Moreover, the Twitter
API does not date the follow edges.\footnote{It is sometimes possible to 
infer when a follow edge was added~\cite{meeder2011we}.}  Due to the long time
interval between crawls of the friend and follower lists, we construct
a static follow graph without time information for all edges crawled
before
February 2018, and daily graphs for the follow edges crawled
during the month.
The \textbf{Quote Graph} is the directed, weighted graph of quote
retweets;  these are tweets that include the URL of another tweet in
their text, along with commentary text.  These are rendered by most
Twitter clients to include a box of the quoted tweet within the box of
the quoting tweet.  A weighted edge $(u_1, u_2, w)$ indicates that
there are $w$ quote-retweets by user $u_1$ that quote tweets posted by
user $u_2$.  As with all other dated relations, we consider the edge
to have the date of the quote, not the original post, and
extract daily aggregates 
for 
all of
February 2018.

The \textbf{Reply Graph} is the directed, weighted graph where an edge
$(u_1, u_2, w)$ indicates that user $u_1$ has posted $w$ tweets that
directly reply to tweets posted by user $u_2$.  Since tweet objects
returned by the Twitter API are dated, this graph is also dynamic, 
and we compute separate reply graphs for each day of February 2018.

The \textbf{Retweet Graph} is the directed, weighted graph where an
edge $(u_1, u_2, w)$ indicates that user $u_1$ has retweeted $w$
tweets originally posted by user $u_2$.  This graph is also dynamic,
as retweets are dated.  Similarly to the Reply Graph, we 
compute separate retweet graphs for each day of February 2018.

\hide{
\spara{Mention Graph.}  The directed, weighted graph of the mention relation as reported by the Twitter
API.  Specifically, tweet objects returned by Twitter include a
\texttt{user\_mentions} field, listing all users whose Twitter screen
name is mentioned in the posted tweet.  These include (i) direct
replies that correspond to the Reply Graph, indirect replies
\emph{e.g.}, when a thread of replies involves multiple users, (ii)
``top-level'' mentions that are not replies, but occur because a user
has explicitly typed the screen name of another user into their tweet
text, and (iii) retweets, as these are automatically counted by
Twitter as mentions.
\reminder{Figure~\ref{} shows the in-degrees, out-degrees etc}
}
 
\hide{ 
\spara{Favorite Graph.} 
The directed, weighted graph where an edge $(u_1, u_2, w)$ indicates
that user $u_1$ has clicked on the ``favorite'' icon for $w$ different
tweets posted by user $u_2$.  Note that whereas all the above graphs
are nearly fully-crawled, due to the rate limiting constraints of the
Twitter API, the crawled favorites are a small sample of the actual.
Moreover, the Twitter API returns a list of users that favorited a
tweet but this information is not dated. Thus, in contrast to the
other dynamic graphs presented above, where each edge is dated to the
creation of the edge, we date the favorite graph based on the date the
original tweet was posted, which is of course before the favorite
relation is created.
\reminder{Figure~\ref{} shows the in-degrees, out-degrees etc}
}

\section{Results}
\label{sec:results}
Our data collection gives us daily graphs of replies, quotes, and retweets. 
But since the Twitter API does not provide information regarding when a {\em
follow} interaction took place, we handle the follow interactions separately
and carefully.  We start with a static snapshot of the follow graph, consisting
of all observations made by our crawler prior to February 2018. 
We then add the implied {\em follow} interactions induced by the
set of nodes involved in either a {\em retweet, quote} or a {\em
reply} interaction that we observed.  Finally, for each day of February,
we also add newly formed follow interactions, if observed by 
our crawler on that day.
Table~\ref{tab1} summarizes our dataset and the pairwise overlap between sets
of nodes and edges appearing in multiple layers, measured
by the {\em overlap coefficient} between two sets 
$X$ and $Y$:   $\frac{|X\cap Y|}{\min{(|X|,|Y|)}}$.
We see that those Twitter users 
involved in retweets overlap the most with all other types of
interactions.

\begin{table}[t]
\caption{Number of users, directed edges and fraction of common
interacting nodes and edges using overlap coefficients (Feb 1--28, 2018)}
\label{tab1}
\centering
  \begin{tabular}{|l||r|r||l|l|l|l||l|l|l|l||}
  \hline
    & \multicolumn{1}{l|}{Nodes} & \multicolumn{1}{l|}{Edges} & \multicolumn{4}{|c|}{Node Overlap Coefficient} & \multicolumn{4}{|c|}{Edge Overlap Coefficient} \\
  \hline
   All types& 1,125,044& 5,000,833 & F & Q & R & RT & F & Q & R & RT \\
  \hline
  Follow(F) &143,453 & 1,082,997 & \cellcolor{orange}    1    &      &      &       & \cellcolor{orange}    1    &      &      &   \\
  Quote(Q)   &   271,824 &   666,820& \cellcolor{orange!44} 0.44 & \cellcolor{orange} 1    &      &     & \cellcolor{orange!17} 0.17 & \cellcolor{orange} 1    &      &    \\
  Reply(R)   &   530,956 & 1,259,970 & \cellcolor{orange!58} 0.58 & \cellcolor{orange!57} 0.57 & \cellcolor{orange}    1    &   & \cellcolor{orange!21} 0.21 & \cellcolor{orange!10} 0.10 & \cellcolor{orange}    1    &   \\
  Retweet(RT) &   762,459 & 3,501,240& \cellcolor{orange!90} 0.90 & \cellcolor{orange!63} 0.63 & \cellcolor{orange!42} 0.42 & \cellcolor{orange} 1 & \cellcolor{orange!85} 0.85 & \cellcolor{orange!23} 0.23 & \cellcolor{orange!19} 0.19 & \cellcolor{orange} 1\\
  \hline
  \end{tabular}
\end{table}

\begin{wrapfigure}{r}{0.46\textwidth}
\centering
\includegraphics[scale=0.39]{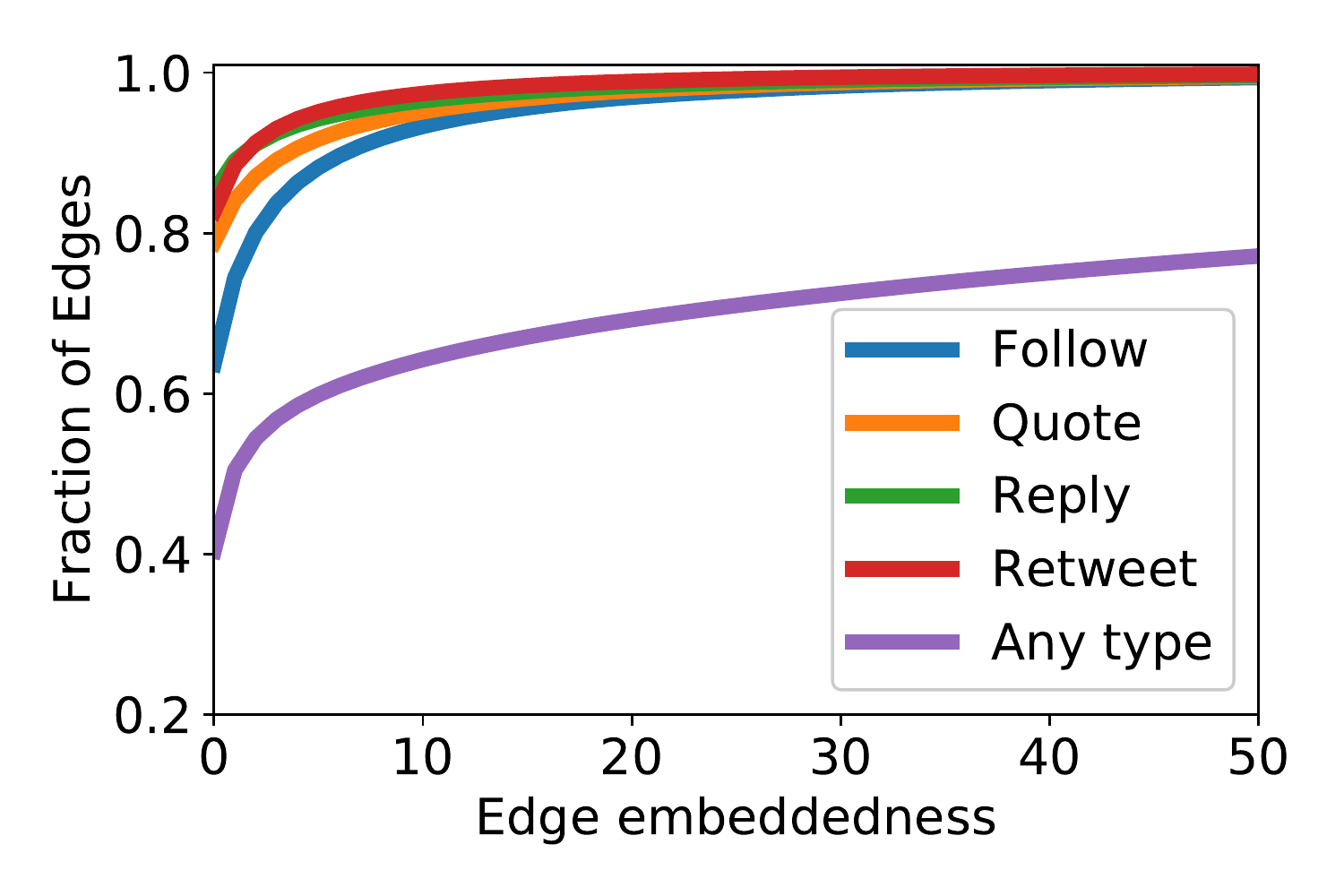}
\caption{Empirical CDF of Edge Embeddedness. Observe that most edges are contained in few triangles across all interaction types.}
\label{cdf} 
\end{wrapfigure}

A different set of important statistics on our layered graph relate 
  to edge embeddedness.
Fig.~\ref{cdf} depicts the empirical CDF of (undirected) edge embeddedness 
  for each of the four layers of the graphs. 
The fifth CDF depicts edge embeddedness for the induced graph that has an 
  undirected edge $(u, v)$ if $u$ and $v$ participated in any type of 
  interaction.
For quotes, replies, and retweets, around 80\% of edges have
  zero embeddedness; for follows, this percentage drops to 63\%, and
  when any type of interaction is considered, only 40\% of edges have
  zero embeddedness.  
Conversely, more than 20\% of edges have any-type
  embeddedness exceeding a value of 50.
Higher embeddedness reflects higher adherence to triadic closure, and 
  we see that the follows interaction has the strongest individual effect. 
The much higher values of embeddedness when considering multiple types of
  interactions additionally demonstrate the promise of using specific 
  triadic features in prediction.

\spara{One hand helping another.}
Our first experiment focuses on the standard link prediction problem,
to test what kind of benefits in terms of prediction accuracy can be
obtained by leveraging additional interactions of other
types. Intuitively, we expect that leveraging information from
additional Twitter layers should strictly improve link prediction. Consider for
instance, a reply interaction. Frequently, replies are correlated with
other interactions, e.g., two users who follow each other may
have first retweeted the same tweet before one replies to the other. 
We verify this intuition experimentally.

As motivated previously in
Section~\ref{sec:intro} (see Figure~\ref{fig:highlight}), we 
quantify the improvements in performance by 
leveraging information from one additional layer 
As in all subsequent experiments, we perform 10-fold
cross validation, and report the {\em classification accuracy}, 
  defined to be the fraction of correct predictions. 
Table~\ref{tab2} reports the average mean
accuracy\footnote{{\it Average} mean accuracy: {\it mean}
refers to the average accuracy over the 10 folds that we obtain for
any single day, and {\it average} refers to the average of the
means over the 28 day period that spans Feb. 2018.} using no extra
layers (first column),  and all extra interaction layers (second column).
The third column shows the relative improvement we obtain when we 
  add all extra layers as part of the training input, 
  clear 
  illustration of the benefits of the
proposed framework. 
Table~\ref{tab2} also shows the standard deviation
from the average mean accuracy over the period of 28 days. It is worth
mentioning that for each day, the accuracies we observe over the
10-folds are well concentrated around their mean. They never exceed
$10^{-3}$ across all days and types of interactions, so we omit
reporting them.

\begin{table}[t]
\caption{\label{tab2} Accuracy and relative improvement from using additional layers}
\resizebox{\textwidth}{!}{
\begin{tabular}{|l|r|r|r|}
\hline
Interaction Type &  Accuracy (one layer) & Accuracy (all layers) & Relative improvement\\
\hline
Follow & 0.732 ($\pm$ 0.003) & 0.796 ($\pm$ 0.002) & 8.83\% ($\pm$ 0.33\%) \\
Quote & 0.629 ($\pm$ 0.004)& 0.728 ($\pm$ 0.014) & 13.57\% ($\pm$ 1.65\%) \\
Reply & 0.585 ($\pm$ 0.003) & 0.770 ($\pm$ 0.002) &  31.62\% ($\pm$ 0.67\%) \\
Retweet & 0.690 ($\pm$ 0.009)& 0.772 ($\pm$ 0.004) &  11.85\% ($\pm$ 1.04\%) \\
\hline
\end{tabular}
}
\end{table}

\spara{Does time affect our predictions?}
We test how our prediction accuracy varies as a function of the number
of days used for crawling Twitter interactions. Specifically, we re-run
our prediction algorithm on an accumulated daily basis, after adding all
new interactions identified by our crawler.
Figure~\ref{daily} plots the retweet and reply prediction accuracy as a
function of time (28 days), which is representative of what we
observe for the rest of interaction types.  Our main observation is
that prediction accuracy remains stable. It is not affected  by
increasing dataset size, nor by seasonal components (e.g., Twitter
activity during weekends vs weekdays), nor by the set of
features we use for training our classifier.   It is worth noticing
that retweet is special in the following way: it is the single type
of interaction where pairwise-type triads perform worse than
retweet-only triads. For all other layers, the $\bigtriangleup$ curve
(single-type triads) lies below the \fbox{$\phantom{o}$} curve
(pairwise triads).
\begin{figure}[t]
  \centering

\minipage{0.7\textwidth}
  \centering
  \includegraphics[width=\linewidth]{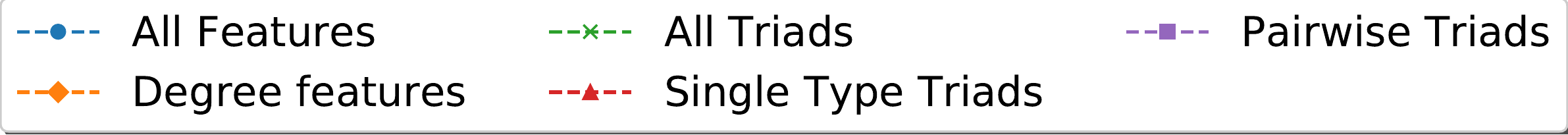}
\endminipage

\medskip

\minipage{0.48\textwidth}
  \centering
  \includegraphics[width=\linewidth,scale=0.95]{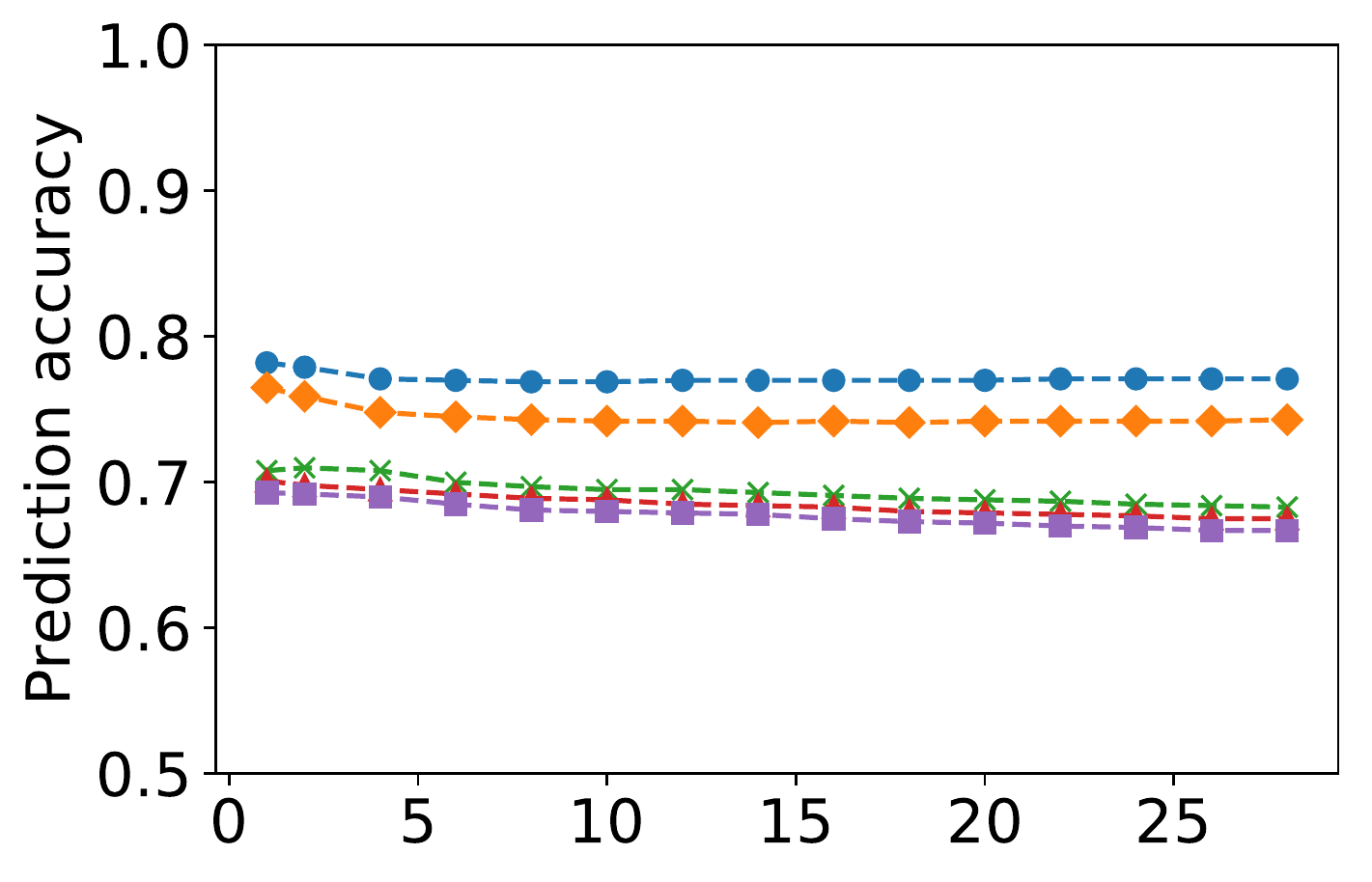}
  (a) Retweet
\endminipage \hfill
\minipage{0.48\textwidth}
  \centering
  \includegraphics[width=\linewidth,scale=0.95]{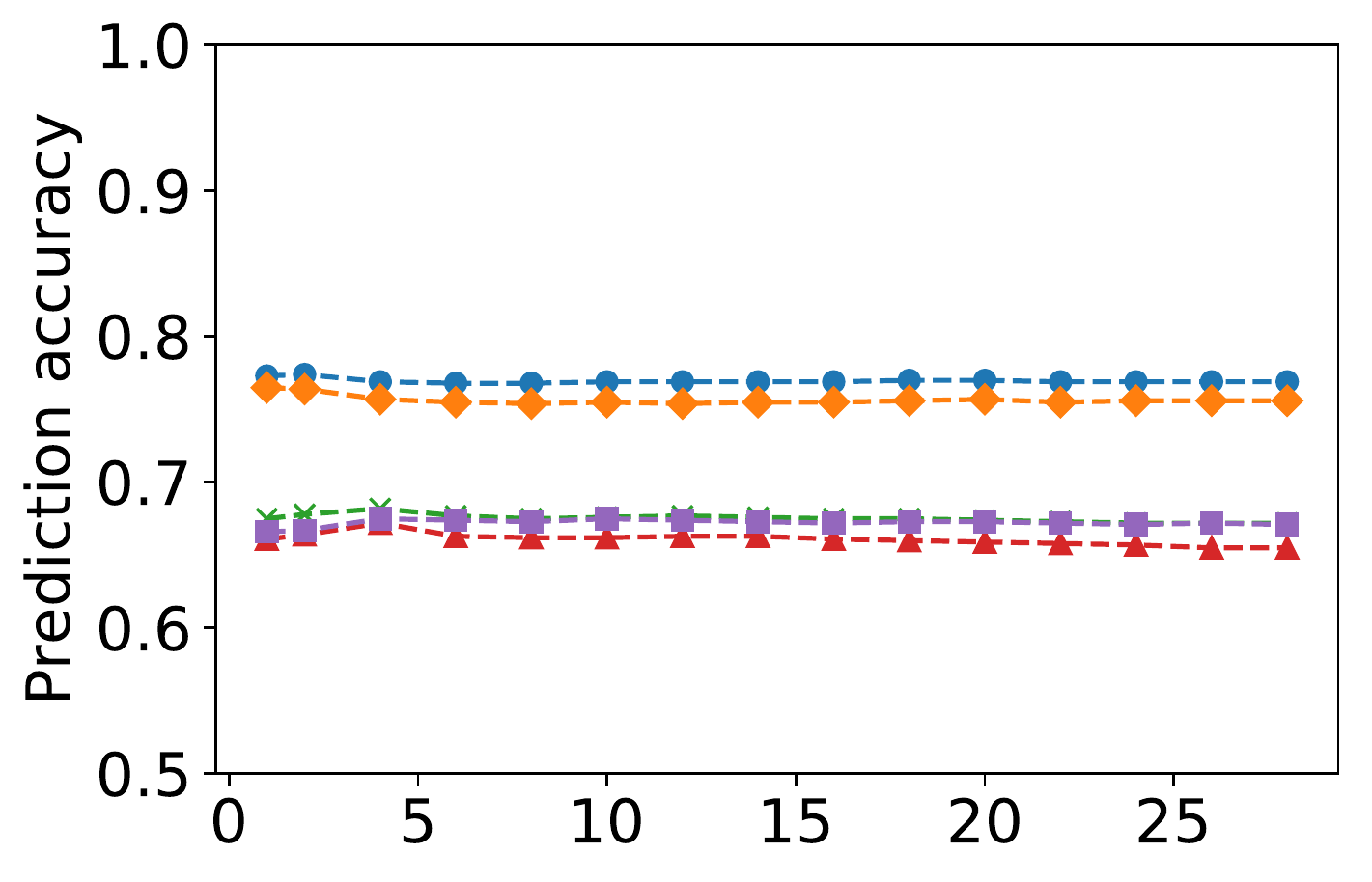}
  (b) Reply
\endminipage \hfill
  \caption{Prediction accuracy for retweets (a) and replies (b) as a function of time (daily scale) using different set of features during training.
}
  \label{daily}
\end{figure}

\spara{Which layer helps most?}
Suppose we want to predict links on a given layer. We now ask, which
of the other layers adds most information to the link prediction classifier? 
This question is important for two reasons. First it provides significant
insights on how Twitter users behave. Secondly, when computational
resources are scarce, then one may want to leverage information from
only one additional layer.  As we can observe from the heat map in
Figure~\ref{fig:highlight}, for all types of interactions, retweets help
most in boosting accuracy.  But, for predicting retweets themselves,
quotes form the most informative layer of interactions. We assume that
this is because quotes are essentially a special type of retweet (even
if they are regarded as different by Twitter), where users
not only retweet the original tweet, but also add their comment.

\hide{ As retweet interactions are the ones that are more prevalent in
our dataset, they form a dense and information-rich graph that we can
exploit. The use of large and dense graphs to infer relationships in
other smaller graphs, has also been used in the work of Rotabi and
others \cite{rotabi2017detecting}, where the follow graph of twitter
was used as a guide to infer strong ties between users in graphs that
were formed from more direct and personal interactions between users,
as the graph of direct messages.  Interestingly, though, when we try
to predict whether a user $u$ has retweeted user $v$, the type of
interaction that we should choose is the ``quote''. Quote is the most
sparse type in our dataset, hence this result could seem surprising.
However, Quotes are very similar to Retweets. Namely, a Quote is a
Retweet, where users not only retweet the original tweet, but they can
also render it inside a box and add their own comment.}

\spara{Overall accuracy and weights of features.}
Table~\ref{overallTable} and Figure~\ref{overallFig} summarize our
results for the classification accuracy when using all data, spanning
the 28 days of February. As we can see, our framework achieves an
accuracy ranging from 71.5\% for the {\em quote} type to roughly 80\%
for the {\em follow} type of interaction.
\hide{ As our prediction task is performed on a balanced dataset, a
baseline for a random predictor is 50\%.}
For all types of features, the degree features are the ones that
achieve the best performance. However, as we will see in the
following, triadic features matter a lot when triads exist in
greater abundance. 

\begin{table}[t]
\caption{Accuracy of prediction using all data}
\label{overallTable}
\centering
\begin{tabular}{|l|r|r|r|r|r|}
\hline
Type &  All & Degree & Single triads& Pairwise triads & All triads\\
\hline
Follow & 0.797 & 0.784 & 0.725 & 0.728 & 0.734  \\
Quote & 0.715 & 0.693 & 0.630 & 0.672 & 0.685 \\
Reply & 0.769 & 0.756 & 0.655 & 0.671 & 0.672 \\
Retweet & 0.771 & 0.743 & 0.675 & 0.667 & 0.683 \\
\hline
\end{tabular}
\end{table}

\begin{figure}[t]
\centering
 \includegraphics[width=0.7\linewidth]{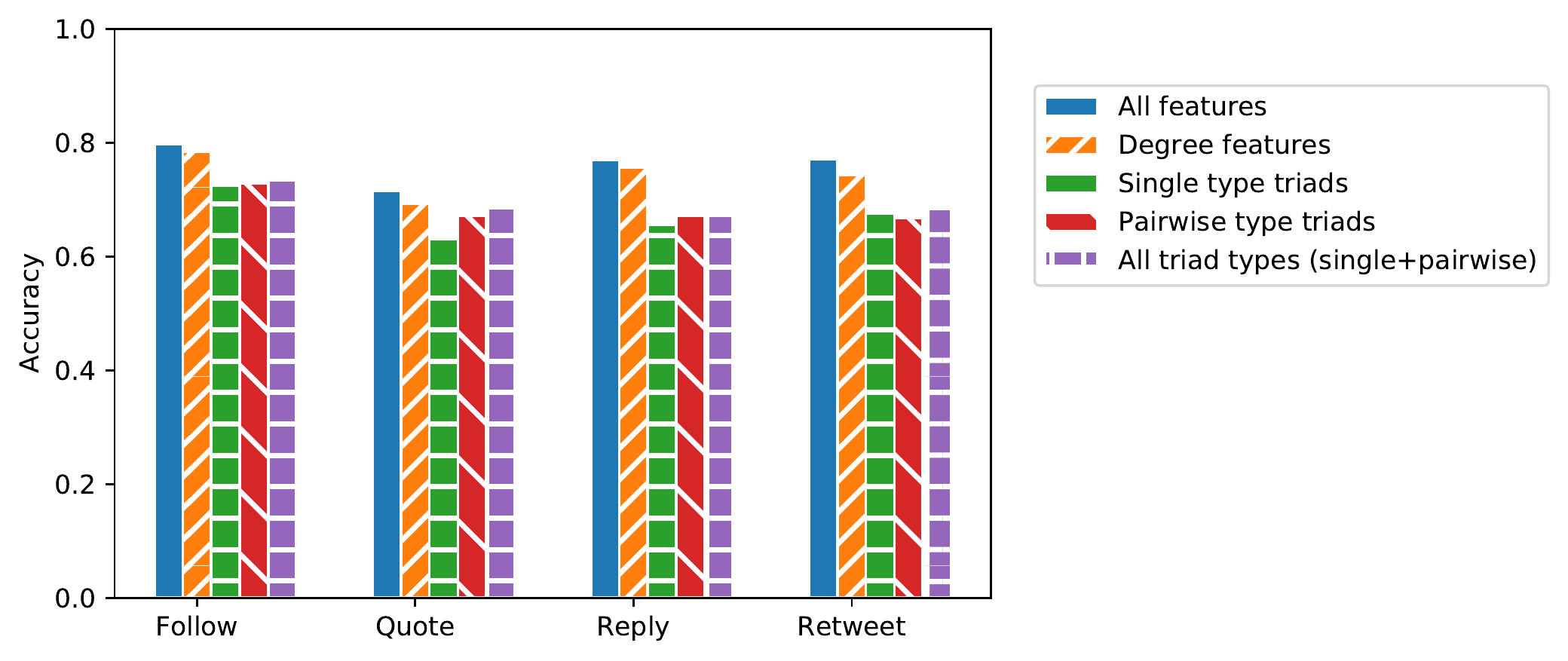}
\caption{Bar chart of accuracy prediction using all data}
\label{overallFig}
\end{figure}

\begin{figure}[h!tb]
    \centering 
\minipage{0.4\textwidth}
  \centering
  \includegraphics[width=\linewidth]{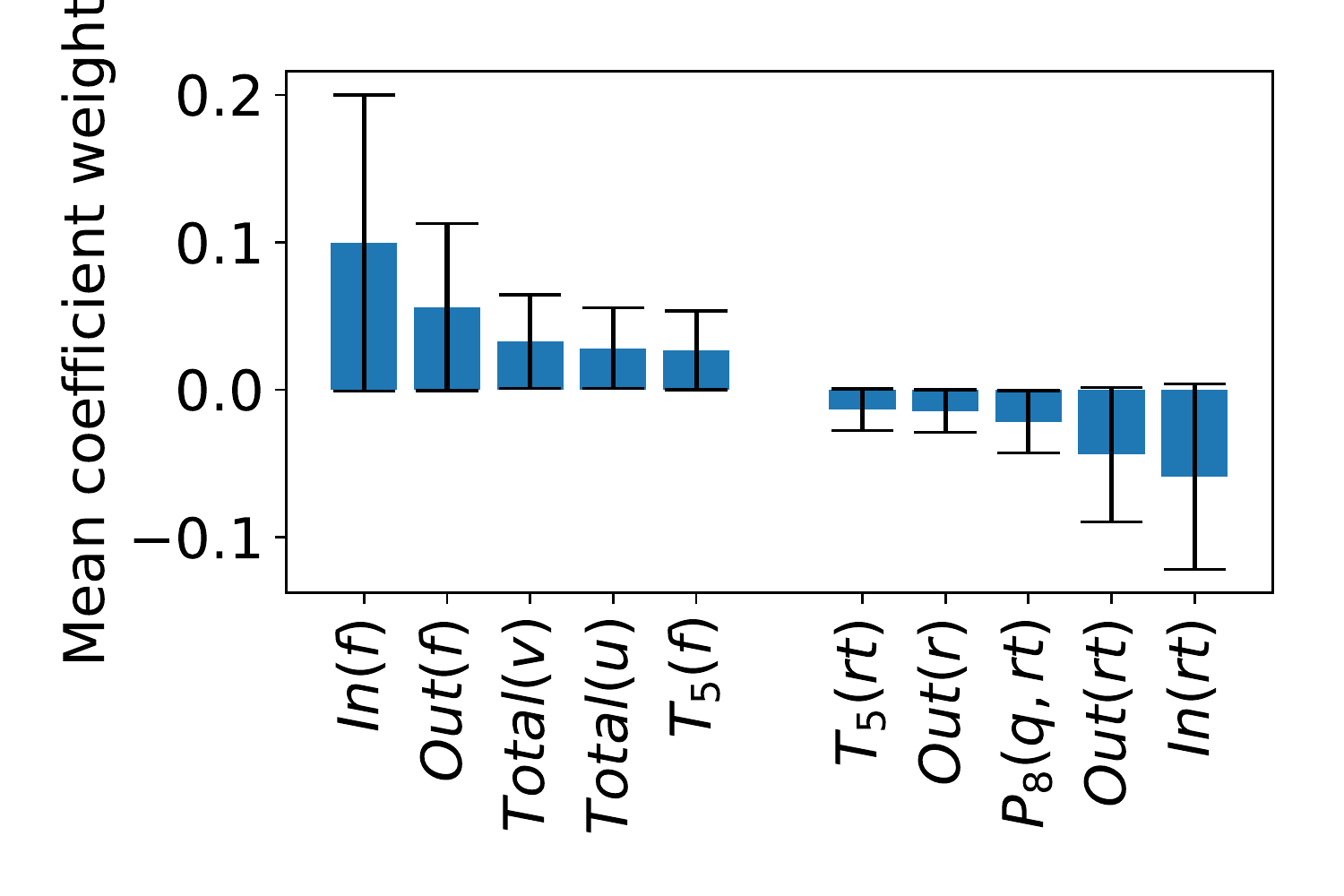}
  (a) Follow
\endminipage \hfil 
\minipage{0.4\textwidth}
  \centering
  \includegraphics[width=\linewidth]{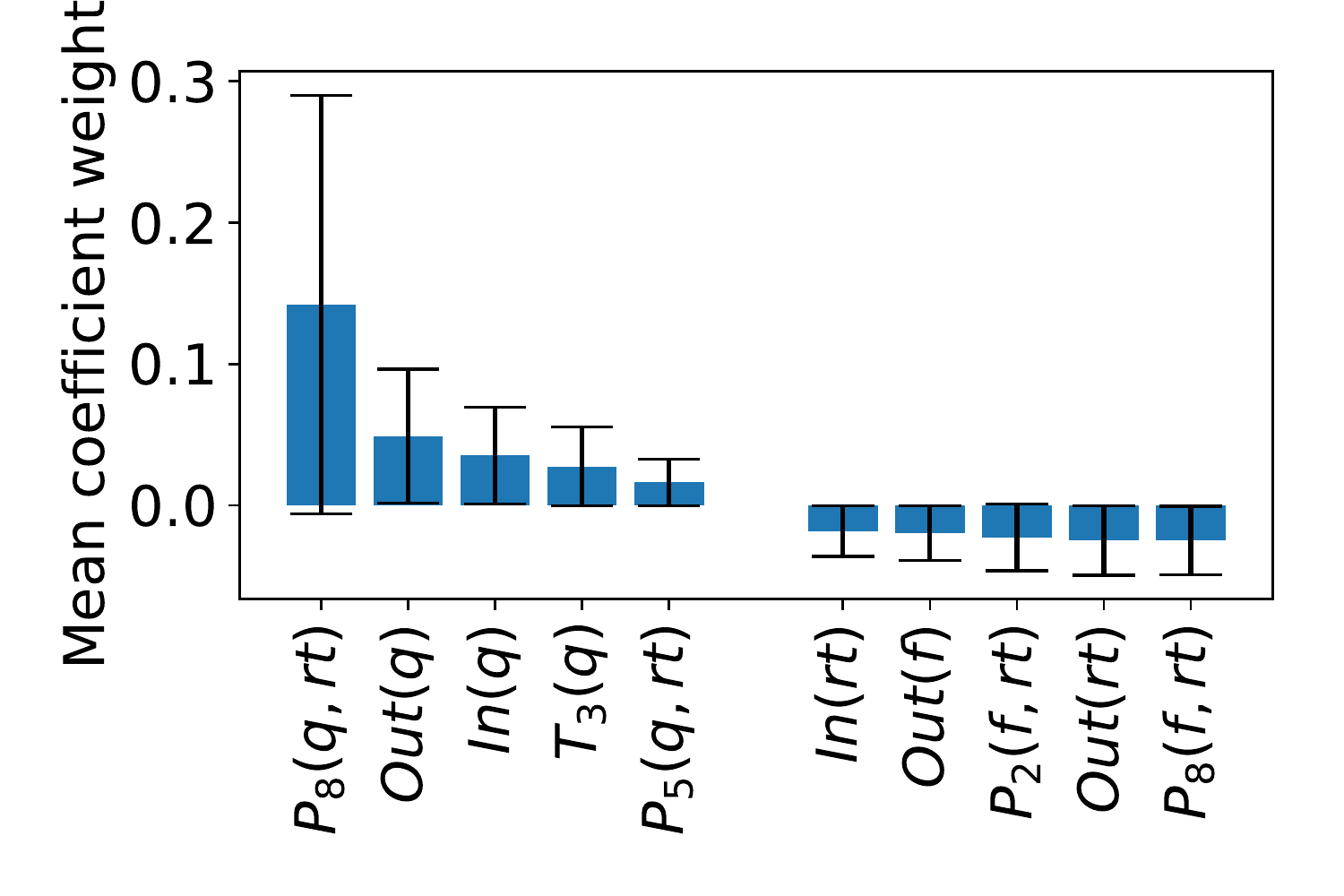}
  (b) Quote
\endminipage \hfil 

\medskip
\minipage{0.4\textwidth}
  \centering
  \includegraphics[width=\linewidth]{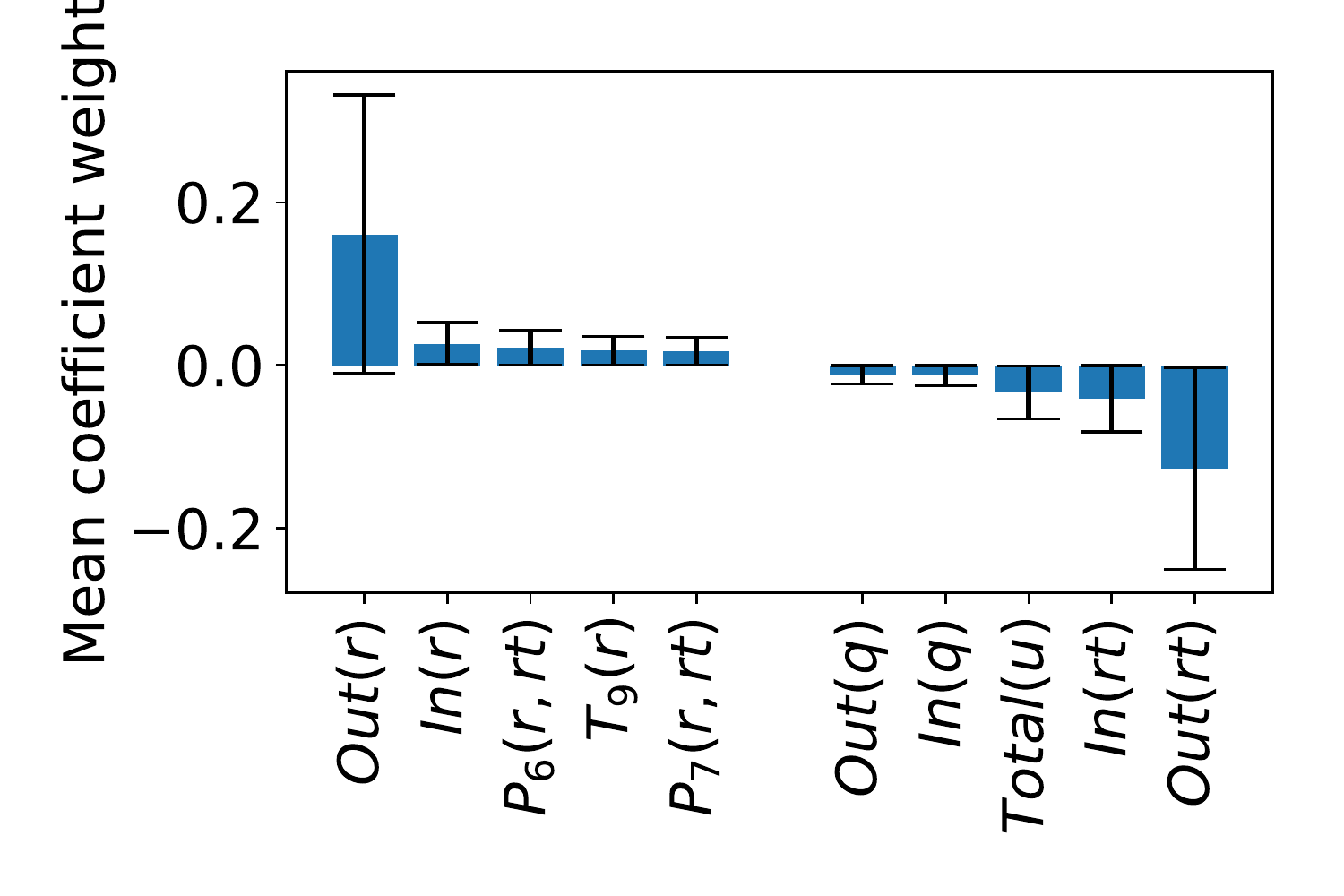}
  (c) Reply
\endminipage \hfil 
\minipage{0.4\textwidth}
  \centering
  \includegraphics[width=\linewidth]{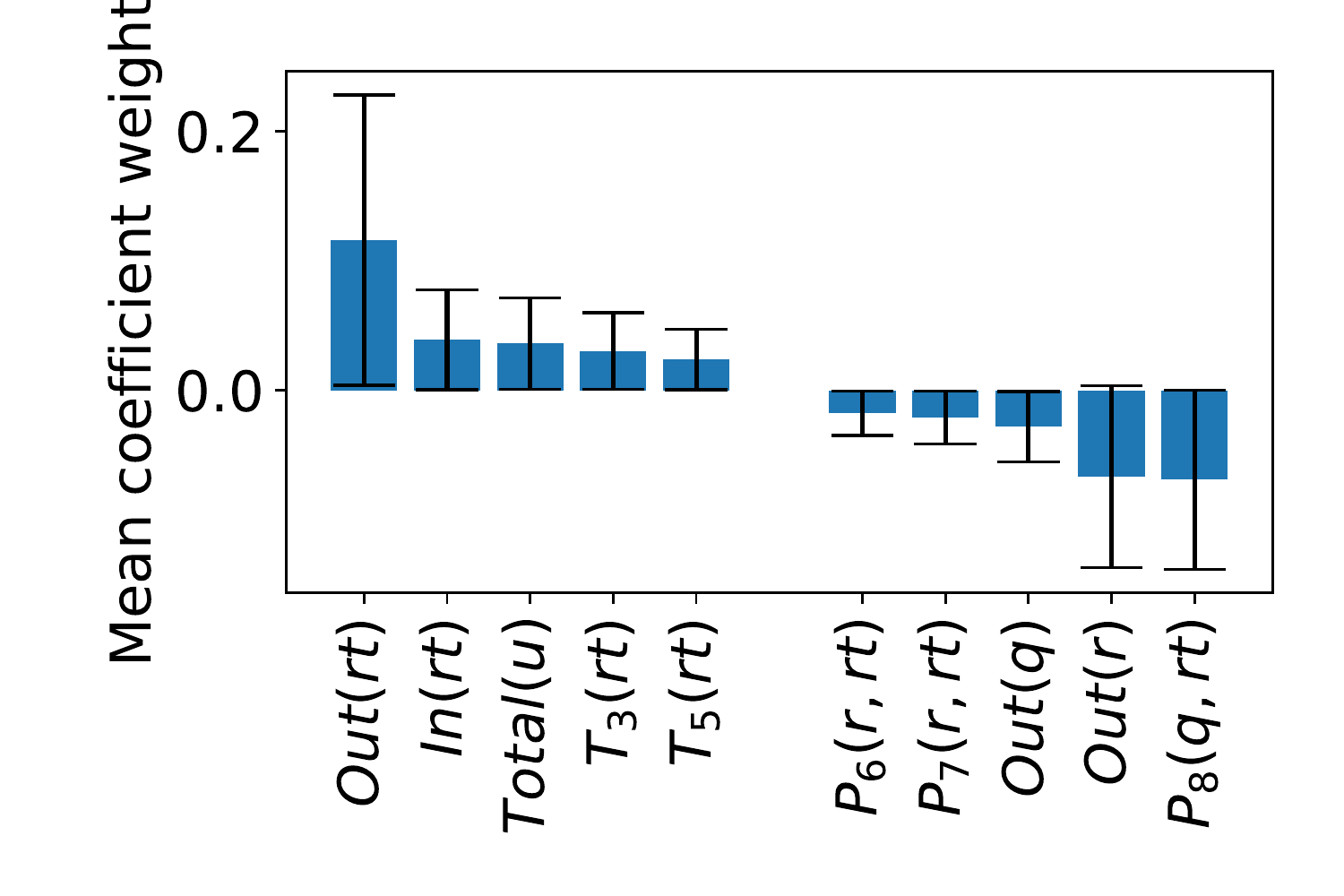}
  (d) Retweet
\endminipage
\caption{Highest (in absolute value) logistic regression coefficient weights}
\label{coef}
\end{figure}

Figure~\ref{coef} displays the features with the highest absolute
value (positive and negative) of their learned coefficient weights. As
the range of each feature differs significantly from one another, we have
standardized them (with a mean value of 0 and standard deviation of 1)
and we also used $\ell_2$-norm regularized logistic regression to obtain
sparse solutions.
From this figure, we see that the degree features are the ones that
play the most important role in predicting interactions. Observe that
that for predicting retweets uses mostly features from the same layer,
and appears to be negatively correlated with other types of
interactions.

Certain triadic features (see Figures~\ref{single_triads} and
~\ref{pairwise_triads} for the id-encoding) ---which are very important
in terms of interpretability, as they can explain patterns of
interactions among users--- play also an important role. 
For follow and
retweet types, we observe that transitive closure, $T_3 (f)$ and $T_3 (rt)$, 
and hierarchy, $T_5 (f)$ and $T_5 (rt)$,  can explain the existence of an edge of this
type. While for quotes, two users that have quoted a common user tend
to also have a retweet relationship between them, $P_8 (q,rt)$. We note that these
are some first findings of our work, as the task of understanding user
behavior on twitter is much broader, and is an interesting open
direction.

\spara{Are degrees or triads more informative features?}
As triadic features constitute a key part of our framework, it is
important to understand when they provide crucial information.
Intuitively, we expect that the prediction accuracy should increase as
the embeddedness of $(u,v)$ increases, simply because these features
make use of an intermediate node $t$ in order to predict an interaction
between $u$ and $v$.   
On the other hand, logistic regression coefficients imply that degree
features are more important than triads.  This naturally brings the
question whether triads or degrees are more important features? The
answer is enlightening, and we explain it in detail in the following: 

\begin{quote}
When a pair of nodes has a large embeddedness value, then triads are
more informative. However, logistic regression coefficients indicate
degree-based features are more important simply because most of the
interactions have zero or very few common neighbors, see
Figure~\ref{cdf}.
\end{quote}

\begin{figure}[t]
    \centering 
\minipage{0.4\textwidth}
  \includegraphics[width=\linewidth]{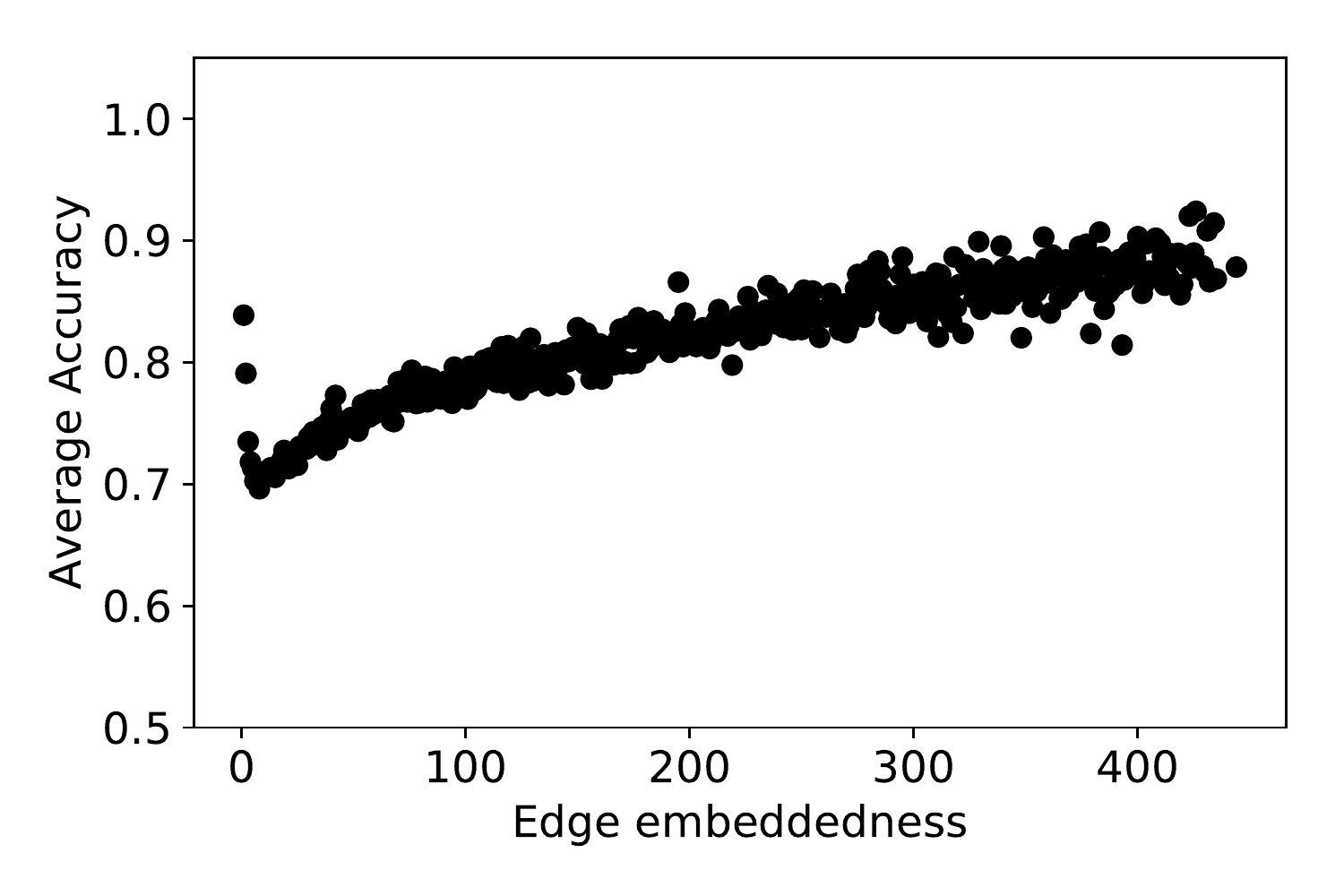}
\centering (a) Follow
\endminipage \hfil 
\minipage{0.4\textwidth}
  \includegraphics[width=\linewidth]{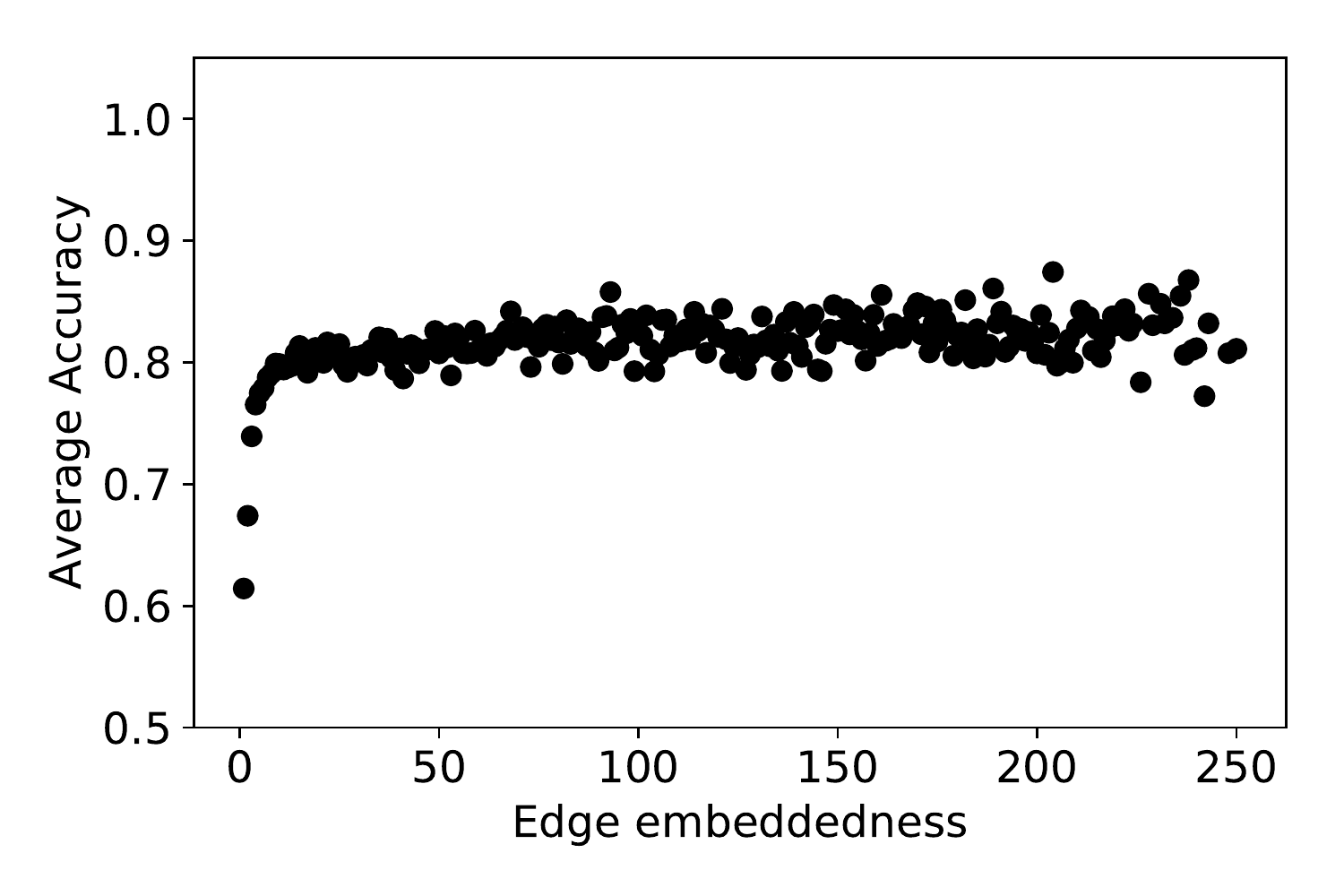}
\centering (b) Quote
\endminipage \hfil 

\medskip
\minipage{0.4\textwidth}
  \includegraphics[width=\linewidth]{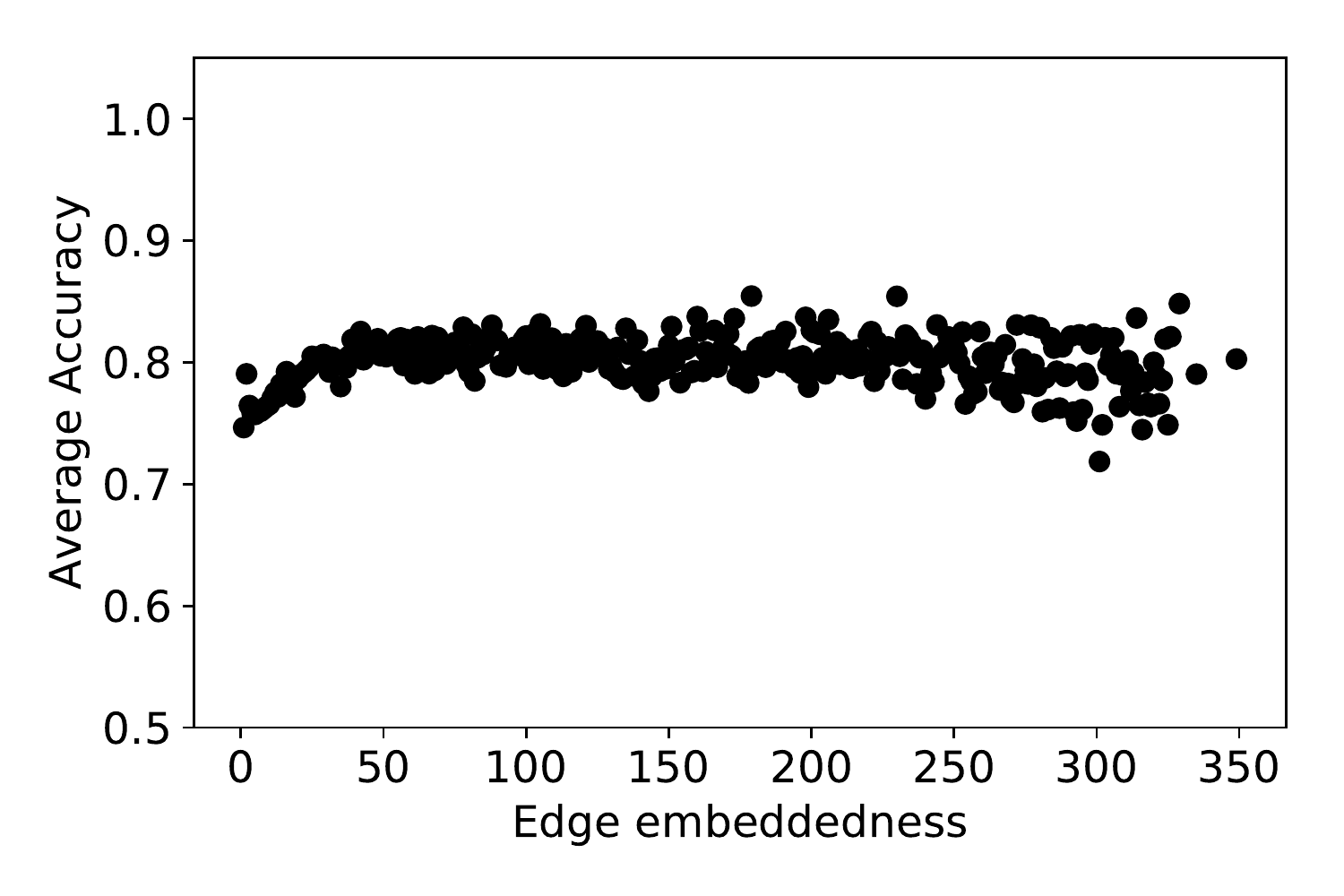}
\centering (c) Reply
\endminipage \hfil 
\minipage{0.4\textwidth}
  \includegraphics[width=\linewidth]{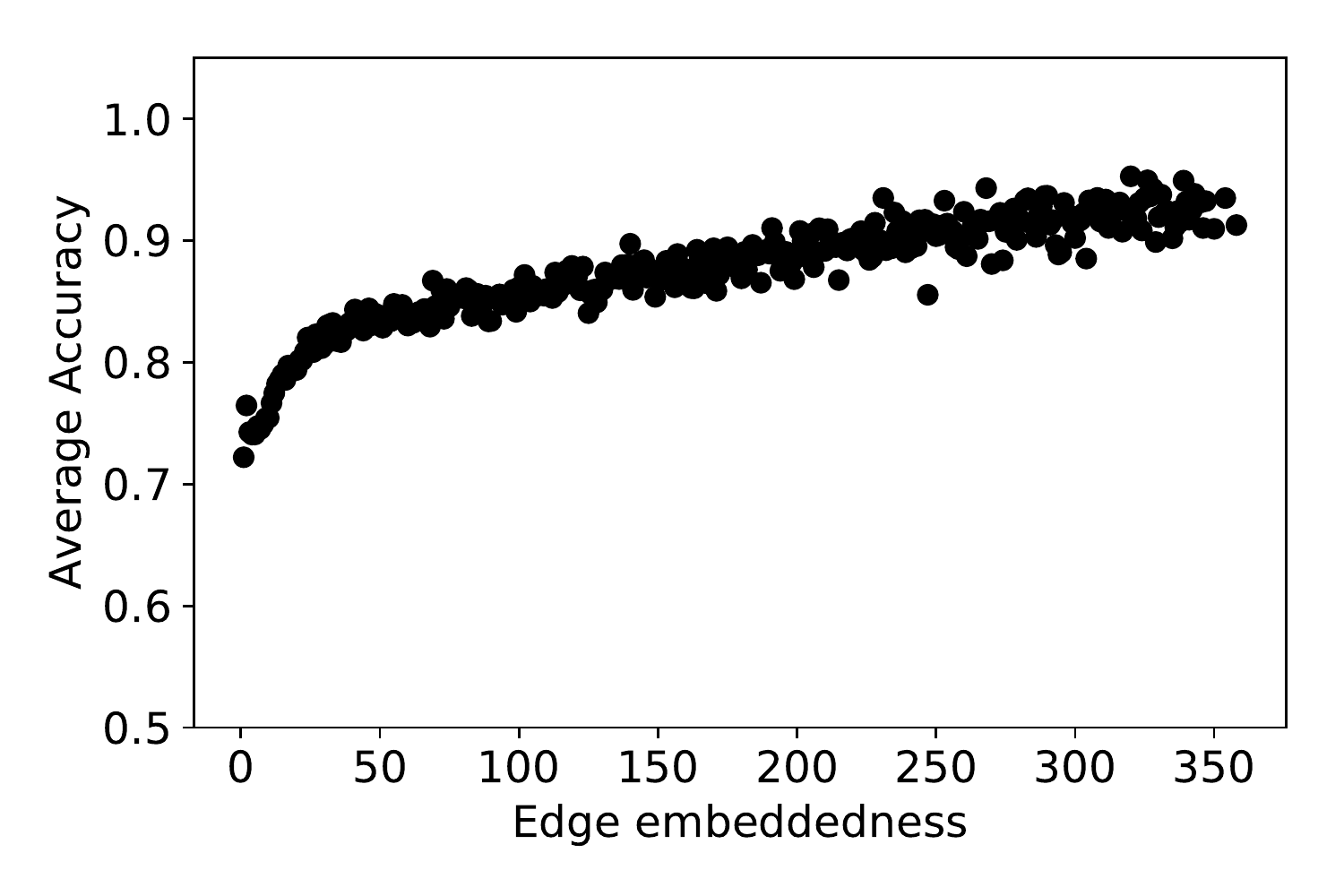}
\centering (d) Retweet
\endminipage
\caption{Average accuracy for edges of certain embeddedness}
\label{accEmb}
\end{figure}

\begin{figure}[t]
    \centering 
\minipage{0.4\textwidth}
  \includegraphics[width=\linewidth,scale=0.9]{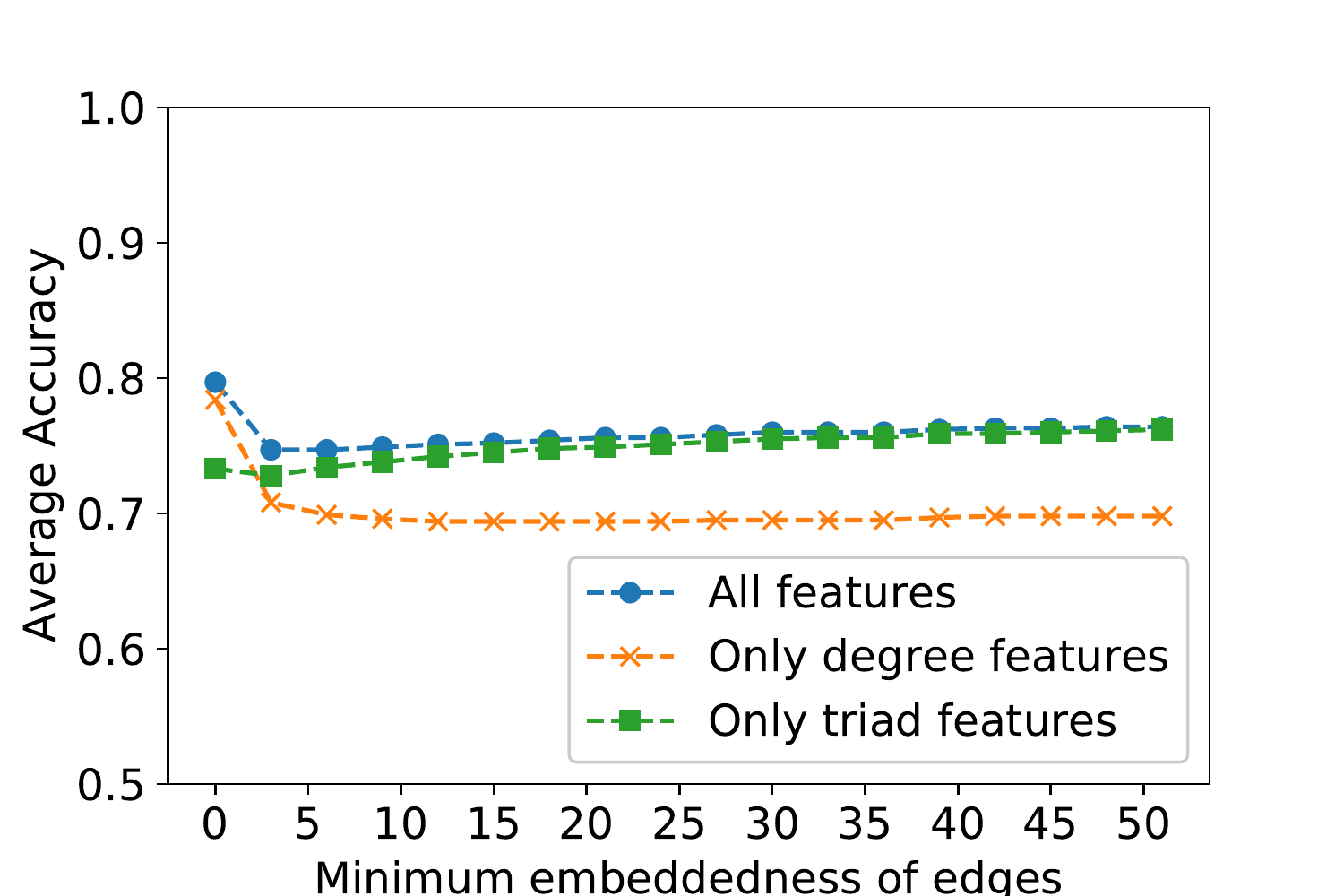}
\centering (a) Follow
\endminipage \hfil 
\minipage{0.4\textwidth}
  \includegraphics[width=\linewidth,scale=0.9]{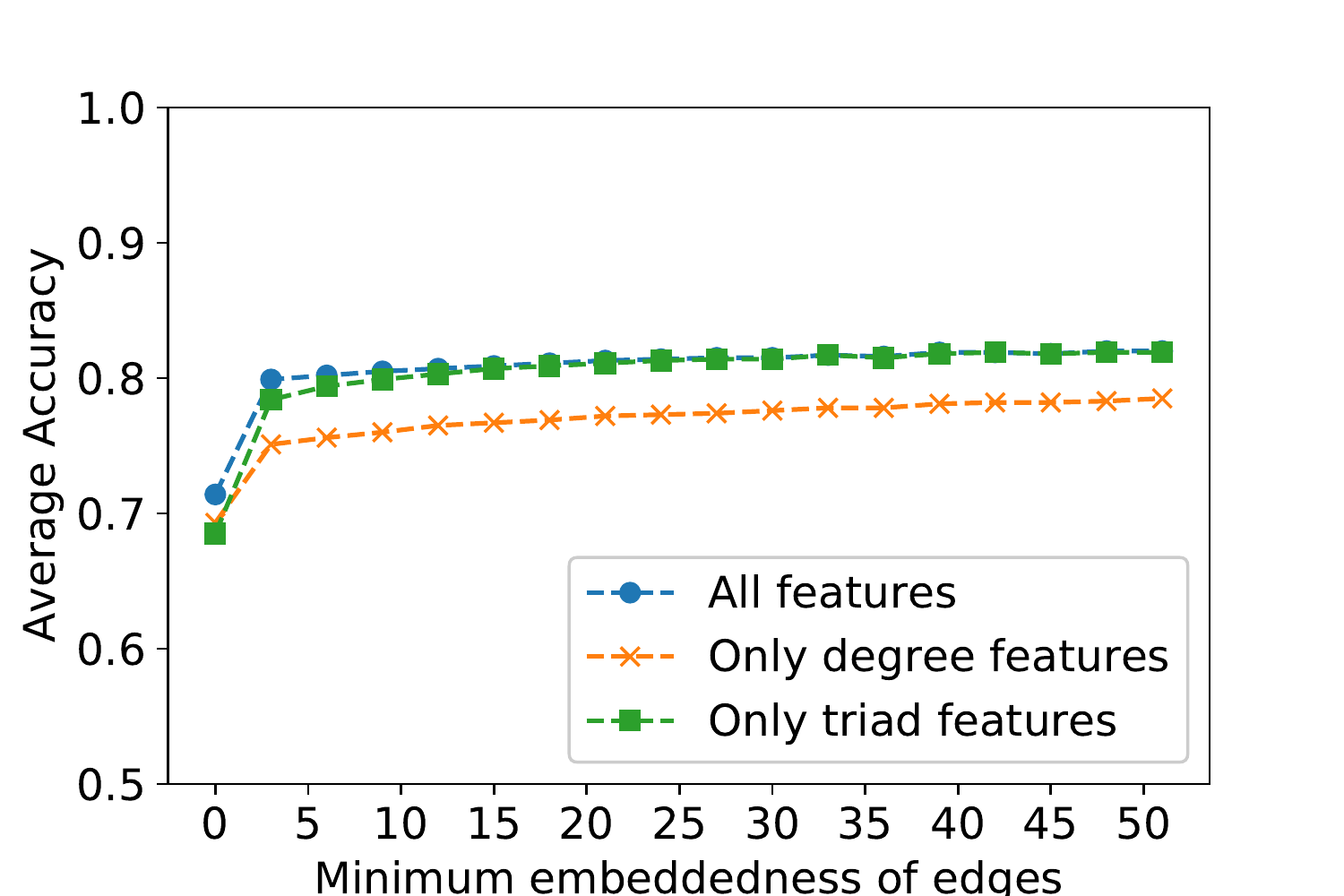}
\centering (b) Quote
\endminipage \hfil 

\medskip
\minipage{0.4\textwidth}
  \includegraphics[width=\linewidth,scale=0.9]{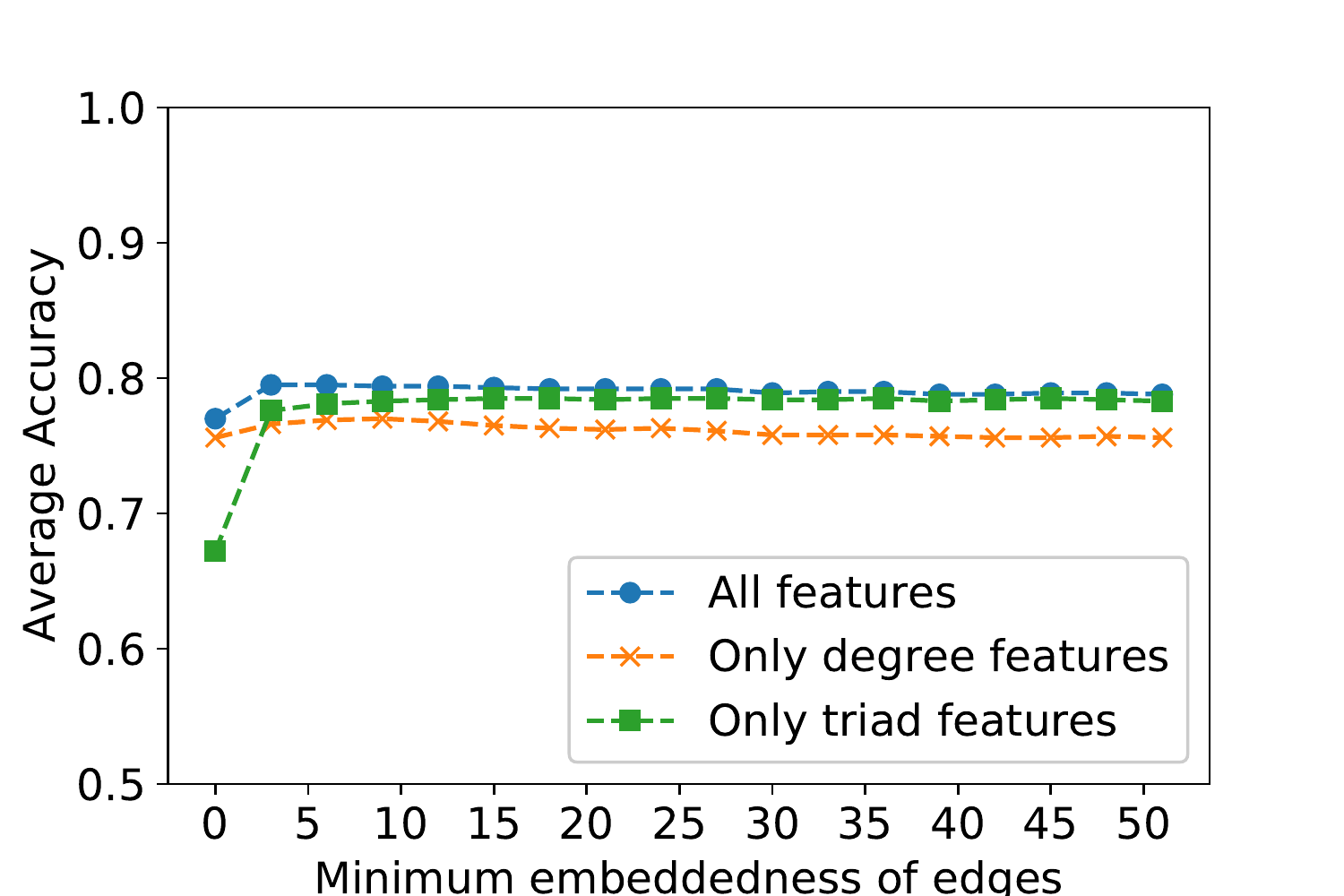}
\centering (c) Reply
\endminipage \hfil 
\minipage{0.4\textwidth}
  \includegraphics[width=\linewidth,scale=0.9]{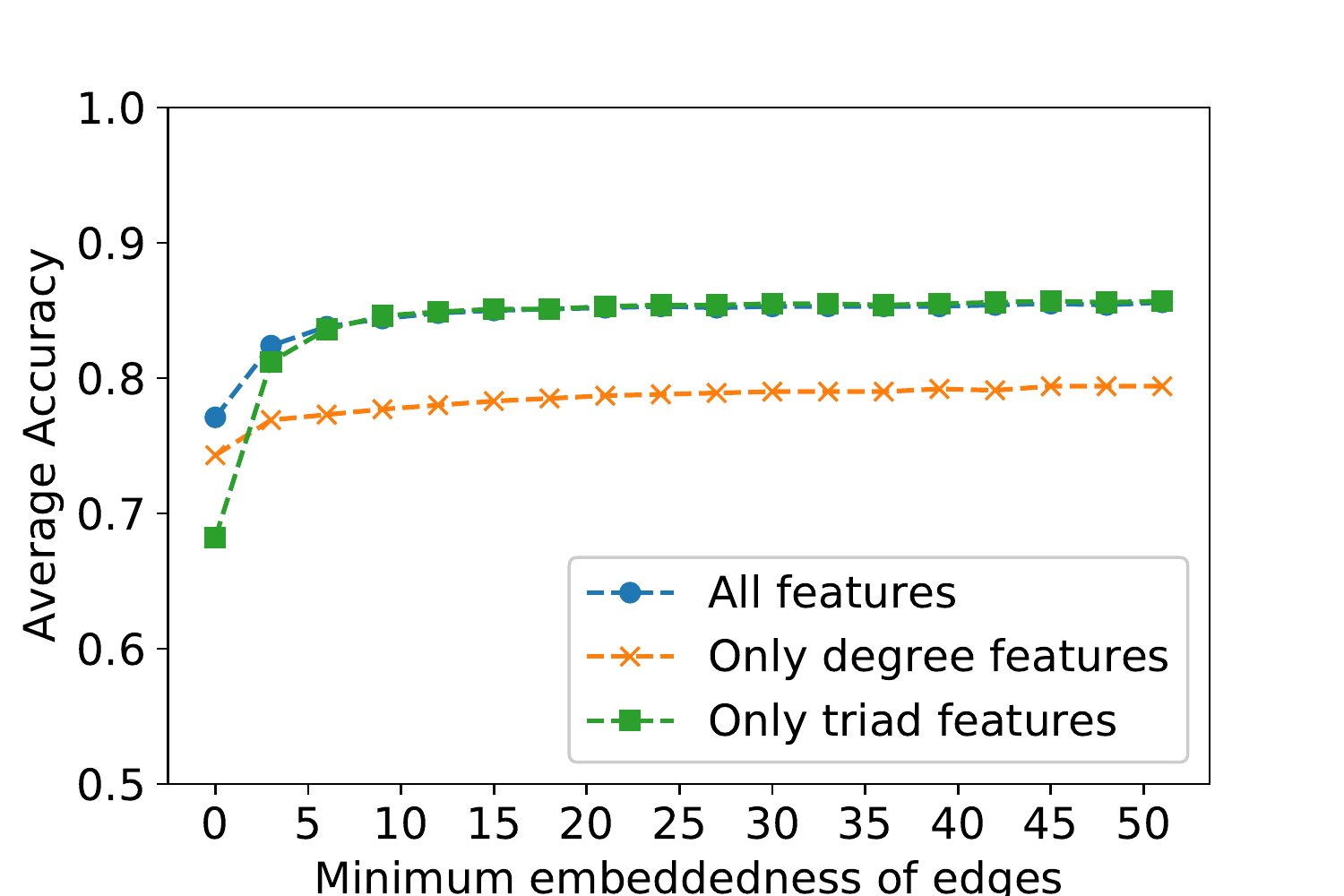}
\centering (d) Retweet
\endminipage
\caption{Prediction accuracy vs.~embeddedness threshold. The figures show the prediction
accuracy when using only edges above a certain embeddedness threshold (x-axis)}
\label{fig9}
\end{figure}

To consider how prediction accuracy varies by embeddedness, 
  we revisit the broadest form of embeddedness:  the number of 
  common neighbors in the undirected graph across all 
  interaction types, i.e., computing embeddedness without regard 
  to directionality of edges or layers. 
We noted that this notion of full embeddedness, as also 
  depicted in Figure~\ref{cdf}, necessarily has higher
  values of embeddedness than any of the layer-specific measures.
Figure~\ref{accEmb}, depict prediction accuracy 
  as a we have grouped the results according to the embeddedness 
  of an edge. 
We find that the task of predicting Quotes and Retweets
  becomes easier as the embeddedness of an edge increases.
Interestingly, for the Follow interaction, we observe that the
accuracy actually decreases for the smallest values of embeddedness
  (with a minimum at 7), which is followed by a 
  steady increase later. 

Our final set of experiments test the efficiency of all of our features,
both degree and triadic, as we restrict attention to subgraphs whose edges
all exceed an embeddedness threshold.
On the $x$-axis of Fig.~\ref{fig9}, we vary
the threshold value for edge embeddedness in order to include 
  it in our dataset, varying it from 0 to 50.
On the $y$-axis, we plot average link prediction accuracy, 
  with three curves for predictions
  using only degree features, only triadic features, and all features, 
  respectively.

In general, prediction accuracy follows observations we made
before. However, a salient and interesting difference is the fact that, while the degree
features were the ones that were leading to higher accuracy previously, now, when every
edge in our dataset meets an embeddedness threshold, the triadic features are those
that become crucial in predicting edges.
Indeed, when the threshold becomes relatively high they tend to explain all variance.
These observations lead us to the following conclusion: while
  degree features are important in predicting edges when the 
  two endpoints have few (or no) neighbors in common, the triadic
  features act complementarily by improving predictions for edges with
higher embeddedness. This agrees with existing findings for
edge sign prediction
\cite{leskovec2010predicting,tsourakakis2017predicting}.

\section{Conclusion}
\label{sec:concl}

\spara{Summary.}
In this work we have studied the link prediction problem on Twitter.
Our approach is based on leveraging a set of different network layers
associated naturally with Twitter activity, namely, the {\em follow,
reply, quote}, and {\em retweet} layers.   Our framework extends the
seminal work of Leskovec, Huttenlocher, and Kleinberg for signed link
prediction \cite{leskovec2010predicting}, and provides significant
insights into how humans behave on Twitter.  Specifically, we find
that by leveraging different layers, results in improving link prediction
accuracy significantly, and that human activity on Twitter is quite 
predictable even for sparse Twitter layers. Among numerous
experiments, we provide a detailed study of which features matter
most for different user profiles, and test aspects of our
framework including sensitivity to time. 

\spara{Open problems.}
Our work opens numerous interesting questions in a range of application 
  domains, including two we consider here: in graph anomaly detection, and in 
  approximate graph inference.
In the first direction, can we use 
existing algorithms~\cite{mitzenmacher2015scalable,tsourakakis2015k}
to locate ``anomalous'' higher-dimensional subgraphs, e.g., $k$-cliques 
for small $k$, or other observed motifs,
and detect subsets of nodes that are dense in these rare subgraphs?
In another direction, we note that rate-limiting of requests to the Twitter 
  API is not specific to our work, but exemplifies a challenge in
  measurement where conducting probes incurs a measurable cost. 
In this setting, maximizing the utility of a set of measurements that 
  is feasible in a cost or time budget becomes paramount, especially
  when there is significant correlation and structure across measurements. 
We view this as especially relevant in scenarios in predictive 
  analytics, where the objective function hinges on 
  prediction accuracy of future queries (such as link predictions) that 
  arrive as an online request stream, not known a priori.

Also, can we use social theories along the lines
of~\cite{leskovec2010predicting} to explain how Twitter users react,
and which modalities of interaction they select?
Finally, are our findings consistent across other subpopulations
of users, e.g., those using either other common languages or 
forming subcommunities around different shared interests?

\bibliographystyle{unsrt}
\bibliography{bibliography}

\end{document}